\documentclass[11pt]{article}

\usepackage[T1]{fontenc}
\usepackage[utf8]{inputenc}
\usepackage[english]{babel}
\usepackage[a4paper,margin=1in]{geometry}
\usepackage{graphicx}
\usepackage{multirow}
\usepackage{amsmath,amssymb,amsfonts}
\usepackage{amsthm}
\usepackage{mathrsfs}
\usepackage{xcolor}
\usepackage{textcomp}
\usepackage{booktabs}
\usepackage{algorithm}
\usepackage{algorithmicx}
\usepackage{algpseudocode}
\usepackage{listings}
\usepackage{caption}
\usepackage{subcaption}
\usepackage{array}
\usepackage{placeins}
\usepackage{makecell}
\usepackage{longtable}
\usepackage{float}
\usepackage{bm}
\usepackage[numbers,sort&compress]{natbib}
\usepackage{authblk}
\usepackage[hidelinks]{hyperref}
\hypersetup{
  pdftitle={Chunk Twice, Embed Once: A Systematic Study of Segmentation and Representation Trade-offs in Chemistry-Aware Retrieval-Augmented Generation},
  pdfauthor={Mahmoud Amiri and Thomas Bocklitz},
  pdfkeywords={retrieval-augmented generation, chemistry retrieval, document chunking, embedding models, ChemQuests, MTEB, scientific question answering}
}

\title{Chunk Twice, Embed Once: A Systematic Study of Segmentation and Representation Trade-offs in Chemistry-Aware Retrieval-Augmented Generation}
\author[1,2]{Mahmoud Amiri\thanks{Email: mahmoud.amiri@uni-jena.de}}
\author[1,2]{Thomas Bocklitz\thanks{Corresponding author: thomas.bocklitz@uni-jena.de}}
\affil[1]{Leibniz Institute of Photonic Technology, Albert-Einstein-Strasse 9, 07745 Jena, Germany}
\affil[2]{Institute of Physical Chemistry and Abbe Center of Photonics, Friedrich Schiller University Jena, Helmholtzweg 4, 07743 Jena, Germany}
\date{}

\begin{document}
\maketitle

\begin{abstract}
The retrieval stage of retrieval-augmented generation (RAG) for scientific question answering depends on how documents are segmented and how chunks are represented in embedding space. This dependence is especially relevant to chemistry texts, which contain dense terminology, symbolic notation, quantitative evidence, and context associated with document structure. However, benchmark-based evidence on the interaction between chunking strategy and embedding model remains limited for chemistry-specific retrieval.

Using \textit{ChemQuests}, a corpus of 952 question-answer pairs from 151 ChemRxiv papers across 17 chemistry subfields, we construct chunk-level, Massive Text Embedding Benchmark (\textsc{MTEB})-compatible retrieval benchmarks for controlled evaluation. We first screen 41 embedding models on the external chemistry retrieval benchmarks {ChemNQRetrieval} and {ChemHotpotQARetrieval} using a geometric-mean metric at rank 10 (Geom@10), which we validate against the full retrieval-metric profile. We then evaluate shortlisted models on ChemQuests-derived tasks across five chunking strategies, seven chunk sizes, and multiple overlap settings.

Embedding choice is associated with the largest observed differences in evidence retrieval, with retrieval-tuned E5, Beijing Academy of Artificial Intelligence General Embedding (BGE), and Nomic models among the strongest overall. Within the evaluated grid, medium-to-large chunks combined with fixed-token, recursive-token, or hierarchical-section chunking provide a practical starting point for the retrieval stage of chemistry-aware RAG. Low overlap was generally favored where overlap variation was evaluated.
\end{abstract}


\section{Introduction}

\begin{figure}[!htbp]
\centering
\includegraphics[width=0.99\textwidth,height=0.75\textheight,keepaspectratio]{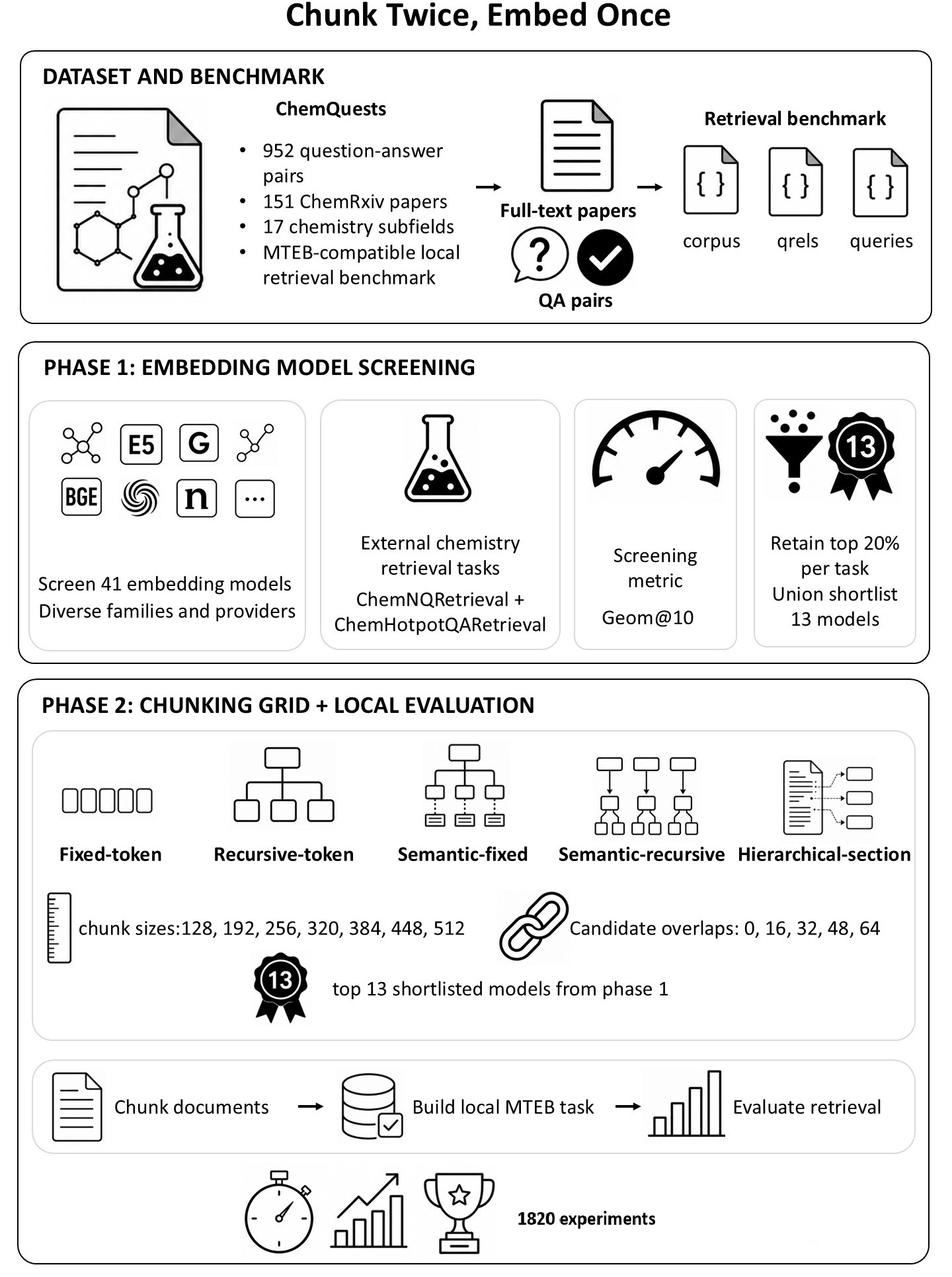}
\caption{Overview of the study design. ChemQuests, a curated chemistry question-answer dataset derived from 151 ChemRxiv papers across 17 chemistry subfields, is converted into a family of \textsc{MTEB}-compatible local retrieval benchmarks, with one benchmark instance generated for each valid retrieval configuration. Each instance consists of corpus, query, and relevance files. The evaluation proceeds in two phases. In Phase 1, 41 embedding models from diverse model families and providers are screened on external chemistry retrieval tasks using the composite metric \(\mathrm{Geom@10}\), and the union of the top 20\% models from each task yields a shortlist of 13 candidate embedding models. In Phase 2, these shortlisted models are evaluated on ChemQuests across a constrained grid of chunking strategies, chunk sizes, and overlap settings. Documents are processed under each chunking configuration, converted into local retrieval tasks, and evaluated to quantify how retrieval configurations affect retrieval quality and computational cost across 1820 experimental configurations.}
\label{fig:study_overview}
\end{figure}

In recent years, the volume of chemistry-related scientific literature has grown rapidly, spanning journal articles, preprints, patents, and technical reports~\cite{ciriminna2024reproducibility}. This expansion has created a pressing need for computational systems that can efficiently retrieve, organize, and synthesize domain-specific knowledge from large document collections~\cite{Lewis2020RetrievalAugmentedGF}. RAG has emerged as a promising paradigm for this purpose by combining information retrieval with language generation, thereby enabling answers that are not only fluent but also grounded in external evidence~\cite{gao2023retrieval}. In chemistry, this grounding is especially important because scientific questions often depend on precise terminology, reaction conditions, measurements, and mechanistic relationships, and errors in retrieved evidence can directly undermine the reliability and interpretability of generated answers~\cite{Skarlinski2024LanguageAAA}.

RAG has been applied in chemistry and related scientific domains~\cite{schwaller2025chemlit, zhong2025benchmarking, lee2025rag, matsumoto2024kragen}, but two retrieval-stage design choices remain insufficiently characterized for chemistry literature: how documents should be segmented into retrievable units and how those units should be represented in embedding space. Chunking determines whether answer-relevant evidence remains within one chunk or is divided across chunk boundaries~\cite{smith2024chunking, yepes2024financial}. Embedding models determine how chemical terminology, entities, and semantic relations are encoded and compared during retrieval~\cite{patil2023survey, kasmaee2024chemteb}. Both choices are often made heuristically, with limited evidence about their interaction in chemistry-specific settings.

The scope of the present work is the retrieval stage of chemistry-aware RAG rather than the full generation pipeline. We evaluate whether retrieval configurations return evidence-bearing chunks under controlled benchmark conditions.

Chunking in RAG is related to linear text segmentation~\cite{ghinassi2024recent, hearst1994multi, choi2000linear, utiyama2001statistical, barzilay2008modeling, misra2011text, riedl2012text, glavavs2016unsupervised, solbiati2021unsupervised, koshorek2018text, lukasik2020text, lo2021transformer, barrow2020joint, somasundaran2020two, gong2022tipster, yu2023improving, fan2023uncovering, jiang2023superdialseg}. Relevant approaches include lexical-cohesion and probabilistic methods, topic models, embedding-based and transformer-based models, supervised and multi-task coherence-aware systems, and zero-shot large language model (LLM) methods. Supplementary Section~1.1 provides further detail. Most of this work has been evaluated in general-domain settings rather than for chemistry-aware retrieval, where section structure, symbolic notation, and tightly coupled local context impose different segmentation requirements.

Chemistry documents contain dense technical language, symbolic notation, structured sections, and evidence that can span adjacent sentences or paragraphs. It remains unclear whether dense retrieval over this literature benefits more from chemistry or scientific pretraining than from retrieval-oriented training objectives. Most prior work evaluates chunking strategies or embedding models in isolation, and common general-domain benchmarks do not represent the lexical and structural properties of chemistry literature~\cite{kwiatkowski2019natural, nguyen2016ms}. Practical guidance for chemistry-aware retrieval systems therefore remains limited.

Accordingly, this study addresses three research questions: (i) which embedding-model families perform most strongly on chemistry-oriented retrieval tasks; (ii) how chunking strategy, chunk size, and overlap affect retrieval over chemistry papers; and (iii) which configurations provide favorable trade-offs between retrieval quality and computational cost.

In this paper, we present a systematic study of chunking-strategy and embedding-model trade-offs in chemistry-focused retrieval. Our evaluation is organized in two stages. In Phase 1, we screen 41 embedding models on two external chemistry-oriented retrieval benchmarks, {ChemNQRetrieval} and {ChemHotpotQARetrieval}, using a compact composite score that summarizes ranking quality and evidence coverage. This stage reduces a broad embedding-model pool to a practical shortlist of strong chemistry-relevant retrieval candidates. In Phase 2, we evaluate the shortlisted models on retrieval tasks derived from \textit{ChemQuests}~\cite{amiri2025ChemQuests} under a constrained grid of chunking configurations spanning five chunking strategies, multiple chunk sizes, and multiple overlap settings. This design makes the joint search over embedding and chunking dimensions computationally feasible and identifies practical retrieval-stage configurations for chemistry-aware RAG systems.

A central contribution of this work is a chemistry-focused retrieval benchmark family derived from \textit{ChemQuests}~\cite{amiri2025ChemQuests}. \textit{ChemQuests} itself is a previously published corpus of question-answer pairs grounded in ChemRxiv. The new contribution in this manuscript is to convert that corpus into a family of chunk-level, \textsc{MTEB}-compatible retrieval benchmarks, where each benchmark instance corresponds to a specific chunking configuration and links chemistry questions to evidence-bearing document chunks. This design provides a controlled in-domain setting for systematic evaluation of segmentation decisions under fixed questions and relevance-labeling procedures, thereby addressing a current lack of retrieval resources tailored to chemistry-specific RAG.

Taken together, this study makes four contributions. First, we provide a systematic empirical evaluation of how embedding-model choice and document chunking jointly affect retrieval over chemistry literature. Second, we derive a family of chunk-level, \textsc{MTEB}-compatible retrieval benchmarks from \textit{ChemQuests}, enabling controlled evaluation of segmentation decisions under fixed chemistry question sets and evidence-labeling procedures. Third, we introduce \(\mathrm{Geom@10}\) as a compact composite screening metric and use correlation and clustering analyses to show that it summarizes complementary ranking-quality and evidence-coverage signals while avoiding over-reliance on mutually redundant retrieval metrics. Fourth, we identify practical retrieval configurations that balance ranking quality, evidence coverage, and computational cost, thereby providing benchmark-supported guidance for the retrieval stage of chemistry-aware RAG pipeline design.

\section{Methods}
This section describes the datasets, chunking strategies, embedding models, evaluation tasks, and pipeline used to assess retrieval over chemistry documents.

\subsection{ChemQuests}
Our study uses \textit{ChemQuests}~\cite{amiri2025ChemQuests}, a curated dataset of 952 question-answer (QA) pairs derived from 151 ChemRxiv papers across 17 chemical subfields. Each QA pair is linked to its source paragraph. As previously reported, the dataset covers organic synthesis, catalysis, materials science, and theoretical chemistry and includes conceptual, mechanistic, applied, and experimental questions. The dataset was constructed using optical character recognition (OCR), QA-pair generation with GPT-4o, and fuzzy string matching to verify alignment with the source text. Ref.~\cite{amiri2025ChemQuests} describes the dataset construction and its application to chemistry-specific natural language processing.

\subsection{External Chemistry Retrieval Tasks}

To assess model transfer before the in-domain ChemQuests chunking study, we benchmarked all models on two established chemistry-focused \textsc{MTEB} retrieval tasks, {ChemHotpotQARetrieval} and {ChemNQRetrieval}. These tasks provide complementary external screening settings, spanning multi-hop chemistry retrieval and more direct open-domain chemistry questions. Dataset details, sources, and access links are provided in Supplementary Section~1.3. We also derive from \textit{ChemQuests}~\cite{amiri2025ChemQuests} a family of chemistry retrieval tasks compatible with \textsc{MTEB}~\cite{muennighoff2022mteb}, with one task generated for each valid chunking configuration.

\subsection{MTEB}
MTEB~\cite{muennighoff2022mteb} is a framework for evaluating text representations across natural language processing (NLP) tasks, including retrieval. In the retrieval evaluation used here, queries and documents are independently encoded as dense vectors using pretrained transformer-based encoder models. The embeddings are normalized to unit length, and cosine similarities are computed by matrix multiplication for exact nearest-neighbor search. For the datasets evaluated here, this procedure provides exact similarity rankings rather than approximate nearest-neighbor results.

\subsection{MTEB Task Creation from ChemQuests}

For each valid chunking configuration, ChemQuests full-text documents were segmented into chunks and exported as a local \textsc{MTEB}-compatible retrieval task. Each task contained {corpus.jsonl} with chunked document units, {queries.jsonl} with ChemQuests questions, and {qrels.jsonl} with binary query relevance (qrel) labels for evidence-bearing chunks. Relevance labels were constructed only within the source document of each question using the supporting evidence snippet and question as lexical anchors, whereas final retrieval evaluation was performed over the full chunked corpus. Details of qrel scoring, thresholding, and the task-construction algorithm are provided in Supplementary Section~1.10 and Supplementary Algorithm~S1.

\subsection{Embedding Models}

We evaluated 41 embedding models spanning retrieval-tuned, scientific, biomedical, chemistry-specific, multilingual, instruction-tuned, and general-purpose encoder families. Full model identifiers and source citations are provided in Supplementary Section~1.6 and Supplementary Table~S1.

\subsection{Chunking Methods}

We evaluated fixed-token, recursive-token, semantic-fixed, semantic-recursive, and hierarchical-section chunking. Fixed-token chunking uses uniform token windows. Recursive-token chunking prefers natural textual separators while enforcing a target chunk size. Semantic-fixed and semantic-recursive first identify semantic regions and then enforce size using fixed or recursive splitting. Hierarchical-section chunking segments within detected article sections while excluding non-retrieval-relevant end matter. Token length was defined with the tokenizer of the embedding model used in each condition, so nominal chunk sizes are model-tokenizer dependent. A visual overview, formal definitions, and implementation assumptions are provided in Supplementary Section~1.4 and Supplementary Figure~S1.

\subsection{Evaluation Metrics}
\label{sec:evaluation_metrics}

Retrieval was evaluated using standard \textsc{MTEB}/Sentence-Transformers retrieval metrics, including \(\mathrm{nDCG@}k\), \(\mathrm{MAP@}k\), \(\mathrm{Recall@}k\), \(\mathrm{Precision@}k\), and \(\mathrm{MRR@}k\), macro-averaged over queries. To summarize ranking quality and coverage in a single score, model retention was based on
\begin{equation}
\mathrm{Geom@10}=\sqrt{\mathrm{nDCG@10}\cdot \mathrm{Recall@10}}.
\label{eq:geom10}
\end{equation}
Here, \(\mathrm{nDCG@10}\) captures ranking quality at rank 10, whereas \(\mathrm{Recall@10}\) captures evidence coverage at rank 10. Their geometric mean rewards models that perform well on both dimensions and penalizes models that achieve high ranking quality at the expense of coverage, or vice versa. This balance is important for chemistry-aware evidence retrieval, where relevant evidence should appear early in the ranked list while still exposing enough supporting material for downstream use. We therefore report \(\mathrm{nDCG@10}\) as the representative ranking-quality metric, \(\mathrm{Recall@10}\) as the representative coverage metric, evaluation time as the principal efficiency indicator, and \(\mathrm{Geom@10}\) as the compact joint screening metric. Detailed metric definitions and formulas are provided in Supplementary Section~1.5.

\subsection{Evaluation Pipeline}

\subsubsection{Phase 1 external screening and model shortlisting}

The evaluation used a two-phase design. Phase 1 was designed to reduce the broad embedding-model search space before the more expensive ChemQuests chunking-grid evaluation. We screened 41 candidate embedding models on two external chemistry-oriented retrieval benchmarks and used \(\mathrm{Geom@10}\) to identify models that jointly performed well in ranking quality and evidence coverage. Phase 1 therefore served as an embedding-model selection step, while the suitability of the compact screening metric was assessed separately using the multimetric validation described below.

To keep Phase 2 computationally tractable while preserving benchmark diversity, we retained the top \(20\%\) of models on each external benchmark, corresponding to the top eight models per dataset. To preserve benchmark-specific retrieval strengths and avoid excluding models that performed strongly on only one retrieval setting, we used the union of the two retained sets rather than selecting models from a single benchmark.

\subsubsection{Phase 2 ChemQuests evaluation and constrained chunking grid}

Phase 2 evaluated the shortlisted embedding models on the ChemQuests-derived retrieval benchmarks under the constrained chunking grid, comprising five chunking strategies, seven chunk sizes ($128, 192, 256, 320, 384, 448, 512$ tokens), and five overlap settings ($0, 16, 32, 48, 64$ tokens). This phase examined the interaction between embedding model and chunking strategy, the effects of chunk size and overlap, and the global trade-off between retrieval quality and evaluation time. Together, these analyses were designed to identify practical retrieval-stage configurations for chemistry-aware RAG.

Not all chunking configurations were evaluated. Let $L$ denote chunk size and $O$ denote overlap. A configuration was considered valid only if it satisfied
\[
O \leq \min(64, 0.3L), \qquad L-O \geq 96, \qquad L+O \leq 512.
\]
These constraints exclude impractical settings with excessive overlap, insufficient effective chunk length, or overly large combined chunk windows. The resulting experiment is therefore a constrained grid search over embedding model, chunking strategy, chunk size, and overlap. Under these constraints, 512-token chunks are valid only at zero overlap. The effect of overlap therefore cannot be evaluated at this chunk size, and 512-token chunks are not included in the overlap-curve analysis.

For each valid configuration, documents were chunked, exported as a local \textsc{MTEB}-compatible retrieval task, and evaluated with the corresponding embedding model using the standard \textsc{MTEB} retrieval pipeline. Full algorithmic details are provided in Supplementary Section~1.11 and Supplementary Algorithms~S2 and~S3.

\subsection{Analysis Procedures}

\noindent\textbf{Metric redundancy and composite-metric validation.}
To assess redundancy among the Phase 1 evaluation criteria, we analyzed pairwise associations among the primary retrieval metrics at rank 10 and evaluation time using Spearman rank correlation and hierarchical clustering. We additionally evaluated whether the compact screening metric \(\mathrm{Geom@10}\) preserved the high-performing model region identified from the full retrieval-metric profile using principal component analysis and clustering of the multimetric model representations. Full methodological details and corresponding analyses are provided in Supplementary Sections~1.8 and~1.9.

\noindent\textbf{Model--chunking aggregation.}
To summarize the interaction between embedding model and chunking strategy, we selected the maximum \(\mathrm{Geom@10}\) over the valid chunk-size and overlap settings for each embedding model--chunking strategy pair (Figure~\ref{fig:model_chunker_heatmap}).

\noindent\textbf{Chunk-size and overlap analyses.}
We analyzed how retrieval effectiveness changes with chunk size and overlap, using \(\mathrm{Geom@10}\) as the summary metric. The overlap analysis was performed at fixed chunk sizes, and the chunk-size analysis was performed at fixed overlap values. In both cases, the plotted curves highlight the top configurations according to normalized area under the curve, together with the mean over all configurations and the mean over the top-performing subset (Supplementary Figures~S11 and~S12).

\noindent\textbf{Performance--evaluation-time analysis.}
To complement the separate analyses of retrieval effectiveness and evaluation time, we examined all evaluated configurations in the two-dimensional space defined by total evaluation time and \(\mathrm{Geom@10}\) as defined in Eq.~\ref{eq:geom10}. In Figure~\ref{fig:global_tradeoff}, each point represents one evaluated configuration, positioned by its evaluation time on the x-axis and its \(\mathrm{Geom@10}\) value on the y-axis. The shaded horizontal region marks the top \(1\%\) of configurations by \(\mathrm{Geom@10}\),defined using the empirical  \(99\)th percentile threshold computed from the full Phase 2 results used to generate Figure~\ref{fig:global_tradeoff}.

\section{Results}


\subsection{Phase 1: Embedding-model screening and metric validation}

Complete 41-model metric profiles are provided in Supplementary Figures~S2 and~S3. The main text reports the leading models in compact tabular form.

\noindent\textbf{Metric redundancy and validation.}
A redundancy analysis of the Phase 1 metrics is provided in Supplementary Section~1.8 and Supplementary Figures~S4 and~S5. Briefly, \(\mathrm{nDCG@10}\), mean average precision at rank 10 (\(\mathrm{MAP@10}\)), and mean reciprocal rank at rank 10 (\(\mathrm{MRR@10}\)) formed a rank-sensitive metric cluster, whereas \(\mathrm{Recall@10}\) and \(\mathrm{Precision@10}\) formed a coverage-oriented cluster. Evaluation time remained weakly associated with retrieval effectiveness.

A multimetric validation showed that \(\mathrm{Geom@10}\) recovered the high-performing model region identified by the full retrieval-metric profile. Details are provided in Supplementary Section~1.9 and Supplementary Figure~S6.

\noindent\textbf{External benchmark performance.}
Table~\ref{tab:phase1_top_models} summarizes the top-ranked models on the two external benchmarks according to \(\mathrm{Geom@10}\). The two benchmarks emphasized different retrieval behaviors and favored partially different model families. \textit{ChemHotpotQARetrieval}, which emphasizes chemistry-focused multi-hop retrieval, ranked Nomic and BGE models highest. In contrast, \textit{ChemNQRetrieval} showed a stronger concentration of E5-family models among the leading configurations. Nevertheless, \textit{nomic-ai/nomic-embed-text-v1.5} appeared in the top tier on both benchmarks, indicating that some retrieval-tuned encoders transferred well across the two chemistry-oriented retrieval settings.

\begin{table}[htbp]
\centering
\caption{Top five embedding models in Phase 1 ranked by \(\mathrm{Geom@10}\) on the two external chemistry retrieval benchmarks.}
\label{tab:phase1_top_models}
\begin{tabular}{llc}
\toprule
Dataset & Model & \(\mathrm{Geom@10}\) \\
\midrule
\multirow{5}{*}{{ChemHotpotQARetrieval}}
& {nomic-ai/nomic-embed-text-v1} & 0.890 \\
& {BAAI/bge-large-en-v1.5} & 0.873 \\
& {nomic-ai/nomic-embed-text-v1.5} & 0.859 \\
& {BAAI/bge-base-en-v1.5} & 0.855 \\
& {intfloat/e5-small} & 0.840 \\
\midrule
\multirow{5}{*}{{ChemNQRetrieval}}
& {intfloat/e5-large} & 0.737 \\
& {intfloat/e5-base} & 0.730 \\
& {nomic-ai/nomic-embed-text-v1.5} & 0.722 \\
& {intfloat/multilingual-e5-base} & 0.713 \\
& {intfloat/multilingual-e5-small} & 0.708 \\
\bottomrule
\end{tabular}
\end{table}

\noindent\textbf{Embedding-model family comparison.}
The main Phase 1 finding was that retrieval-tuned embedding models outperformed domain-pretrained scientific, biomedical, and chemistry-focused encoders on these external retrieval benchmarks. Within the evaluated model set, models such as \textit{SciBERT}~\cite{beltagy2019scibert}, \textit{SPECTER}~\cite{cohan2020specter, allenai2020specter}, \textit{MatSciBERT}~\cite{matscibert2021, gupta_matscibert_2022}, \textit{BiomedBERT}~\cite{gu2021domain, microsoft2021biomedbert}, \textit{ChemBERT}~\cite{recobo2023chembert}, and \textit{ChemBERTa}~\cite{chemberta, deepchem2023chemberta77m} remained below the strongest observed E5-, BGE-, and Nomic-based encoders across the principal retrieval metrics. Within this model set and these benchmarks, domain pretraining alone was not associated with the strongest retrieval results. Models explicitly tuned for retrieval were generally more competitive.

\noindent\textbf{Model shortlist.}
The predefined top-20\% retention procedure yielded a 13-model shortlist, with only three models appearing in both top-eight sets, indicating that the union captured complementary retrieval strengths rather than duplicating a single ranking. The full shortlist is provided in Supplementary Table~S2.

\subsection{Phase 2: ChemQuests chunking-grid evaluation}

\subsubsection{Embedding model--chunking strategy interaction}

\begin{figure}[!htbp]
    \centering
    \includegraphics[width=\textwidth,height=0.75\textheight,keepaspectratio]{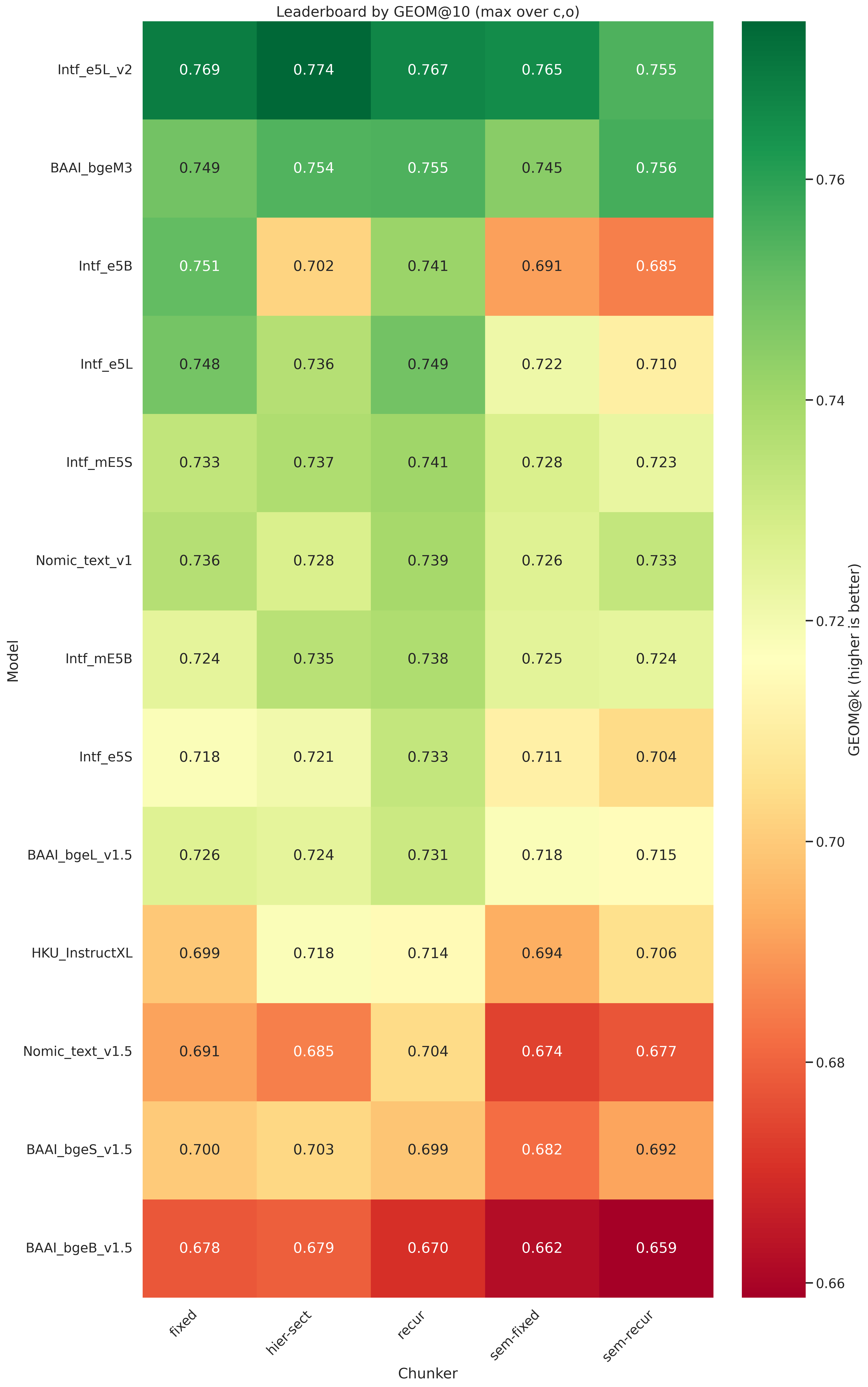}
    \caption{Embedding model--chunking interaction in Phase 2. Heatmap of maximum \(\mathrm{Geom@10}\) across embedding model--chunking strategy pairs, with each cell showing the maximum over valid chunk sizes and overlaps. Retrieval-tuned models, especially E5-family models, generally perform best, with \texttt{intfloat/e5-large-v2} among the top performers. An exception is semantic-recursive chunking, where \texttt{BAAI/bge-m3} achieves the highest score. Across the evaluated configurations, differences among embedding models are larger than differences among chunking strategies.}
    \label{fig:model_chunker_heatmap}
\end{figure}

Figure~\ref{fig:model_chunker_heatmap} shows larger differences in retrieval effectiveness among embedding models than among chunking strategies after aggregation over chunk sizes and overlap settings.

The highest-performing portion of Figure~\ref{fig:model_chunker_heatmap} is dominated by retrieval-oriented encoders, especially E5-family models. In particular, {intfloat/e5-large-v2} attains the highest \(\mathrm{Geom@10}\) for most chunking strategies, including fixed-token, recursive-token, hierarchical-section, and semantic-fixed chunking, whereas {BAAI/bge-m3} attains the highest score for semantic-recursive chunking. Although model ordering varies somewhat across chunking strategies, fixed-token, recursive-token, and hierarchical-section chunking remain consistently competitive.

\subsubsection{Effects of chunk size}

Chunk size shows a clear effect (Supplementary Figure~S12). Retrieval performance generally improves as chunk size increases from 128 tokens toward medium-to-large chunk sizes, after which the gains tend to level off. Small chunks, especially 128-token chunks, underperform across most overlap settings. Performance remains strong in the larger-size regime, with 448-token chunks competitive across the nonzero-overlap curves. For 512-token chunks, zero overlap was the only valid setting under the constrained grid. These configurations performed strongly at zero overlap, but the effect of overlap cannot be evaluated at this chunk size.

\subsubsection{Effects of overlap}

The effect of overlap is generally neutral to negative, with at most small gains in a limited subset of configurations (Supplementary Figure~S11). Across most fixed chunk sizes included in the overlap-curve analysis, the mean \(\mathrm{Geom@10}\) over all configurations decreases as overlap increases. The same tendency is visible for many of the top configurations, although not perfectly monotonically in every case. Several high-performing model--chunking strategy combinations show small local optima at nonzero overlap, especially around 16 or 32 tokens. Overall, large overlap is rarely associated with the highest values in the evaluated grid. For 512-token chunks, zero overlap was the only valid setting, so the effect of overlap cannot be assessed at this chunk size.

The chunk-size and overlap analyses place the highest observed \(\mathrm{Geom@10}\) values in configurations with relatively large chunks and low overlap.

\subsubsection{Global performance--evaluation-time trade-off}

Detailed evaluation-time analyses show that benchmark construction contributes little to the total cost, whereas most runtime variation comes from document segmentation and retrieval evaluation. Encoder choice is associated with the largest differences in total evaluation time, with semantic chunking generally slower than fixed-token, recursive-token, and hierarchical-section chunking. Chunk size and overlap have smaller secondary effects. Supplementary Section~1.12 and Supplementary Figures~S7--S9 provide the full breakdown.

\begin{figure}[!htbp]
    \centering
    \includegraphics[width=\textwidth]{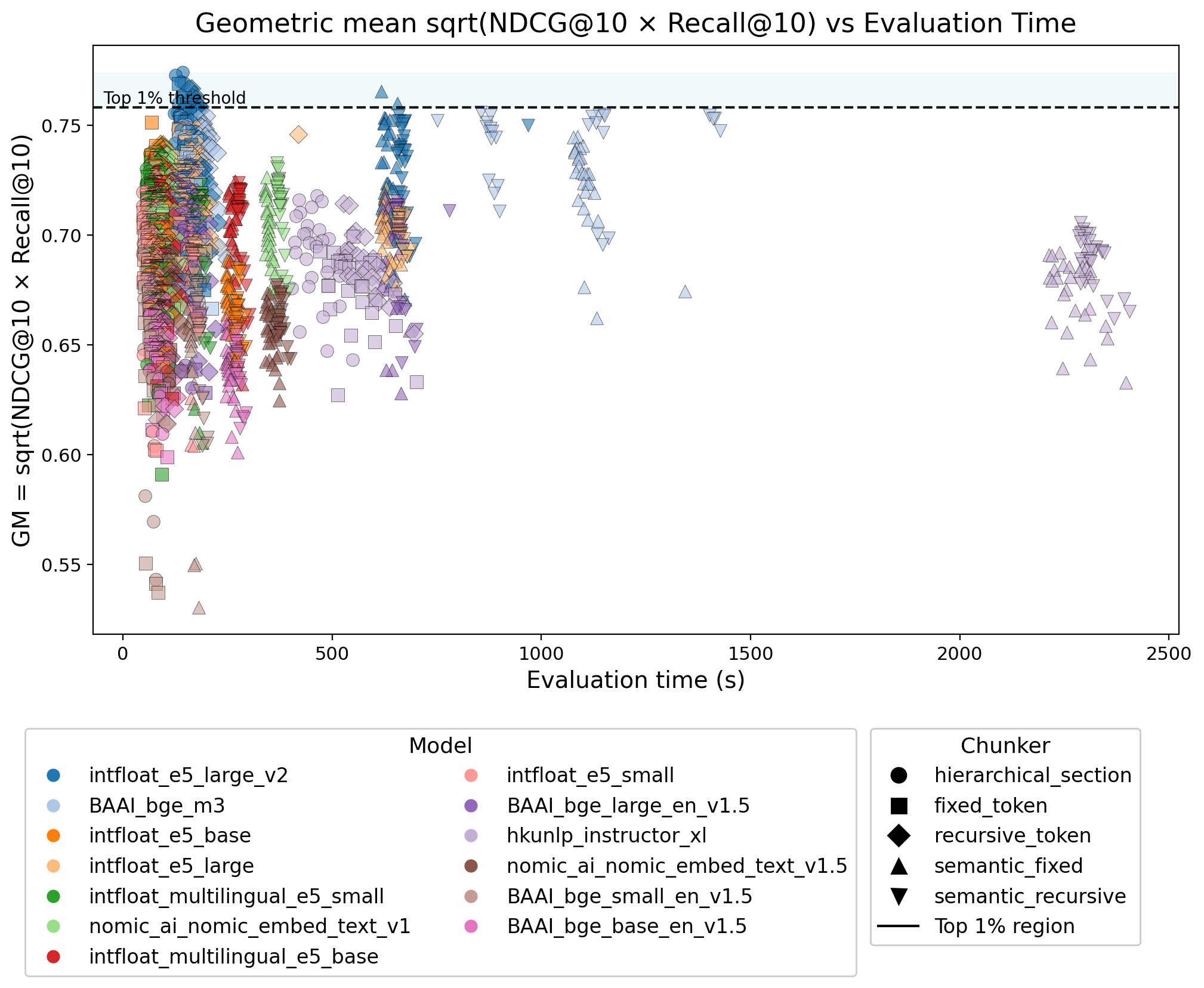}
    \caption{Global relationship between retrieval quality and evaluation time across all Phase 2 configurations. Each point corresponds to one evaluated configuration, plotted by total evaluation time and \(\mathrm{Geom@10}\) as defined in Eq.~\ref{eq:geom10}. The shaded horizontal band marks the top \(1\%\) of configurations by \(\mathrm{Geom@10}\), i.e., all runs above the empirical \(99\)th percentile threshold of the composite score. High-scoring configurations span a range of evaluation times. Some top-performing runs are comparatively fast, whereas others are substantially slower. Evaluation time is therefore only weakly associated with retrieval effectiveness in the evaluated configurations.}
    \label{fig:global_tradeoff}
\end{figure}

Figure~\ref{fig:global_tradeoff} shows that the highest-observed quality configurations are concentrated near the top of the plot, but they are not restricted to a single narrow evaluation-time value. Several top-performing runs occur at relatively modest evaluation times, while others are substantially slower. Thus, the figure does not indicate a simple monotonic relationship between evaluation time and retrieval quality. Increasing evaluation time does not, by itself, guarantee higher \(\mathrm{Geom@10}\), and several comparatively expensive configurations remain only average in retrieval effectiveness.

At the same time, all configurations in the top-\(1\%\) region use {intfloat/e5-large-v2} with chunk sizes of 384--512 tokens, and most use low overlap of at most 32 tokens. Fixed-token, recursive-token, and hierarchical-section chunking appear repeatedly in this region, whereas semantic chunking strategies are less frequent and often require more evaluation time elsewhere in the analysis.

A supplementary multimetric comparison of the top-performing configurations confirms that the leading runs remain strong across the principal retrieval metrics; all use intfloat/e5-large-v2 with large chunks and low overlap (Supplementary Figure~S10).

\section{Discussion}

Within the evaluated configurations, embedding-model choice was associated with larger differences in retrieval quality than chunking strategy. Across the external chemistry benchmarks and the ChemQuests-derived retrieval tasks, the evaluated retrieval-tuned models from the E5, BGE, and Nomic families generally performed better than the evaluated scientific, biomedical, and chemistry-pretrained encoders. These results suggest that retrieval-oriented training objectives may contribute more to dense evidence retrieval than domain pretraining alone within this model set. Domain-specific encoders may still be useful for other chemistry NLP tasks, but in this retrieval-stage evaluation they did not match the strongest retrieval-aligned alternatives.

Larger chunks with little or no overlap were generally preferable within the evaluated grid. Retrieval performance improved from small chunks toward medium-to-large chunk sizes, with gains tending to level off in the larger-size regime. One possible explanation is that very small chunks divide evidence such as reaction conditions, quantitative results, material properties, and mechanistic explanations across retrieval units. Overlap had a weaker and often negative association with retrieval quality. Although some configurations showed small gains at low nonzero overlap, large overlap was rarely associated with the best results and increased redundancy in the indexed corpus.

The comparison of chunking strategies indicates that more complex segmentation does not automatically improve chemistry retrieval. Fixed-token, recursive-token, and hierarchical-section chunking were consistently competitive, while the evaluated semantic chunking strategies produced mixed results and added computational cost. This may be because chemistry evidence often depends on local technical continuity rather than broad topic-level coherence. Simple large-window chunking can preserve adjacent experimental details and interpretations without relying on potentially noisy semantic boundary detection. Hierarchical-section chunking remains attractive when reliable article structure is available because it respects scientific document organization while avoiding irrelevant end matter.

For similar chemistry retrieval tasks, a practical starting configuration would pair an encoder from the E5, BGE, or Nomic family with fixed-token, recursive-token, or hierarchical-section chunking, using large chunks and low or zero overlap. Semantic chunking, very small chunks, and large overlaps should be used cautiously unless the target corpus or downstream task gives a specific reason to prefer them. The global quality--time analysis shows that higher computational cost does not guarantee better retrieval; retrieval quality and runtime should therefore be considered jointly rather than assuming a monotonic cost--performance trade-off.

Several limitations should be considered. The ChemQuests-derived relevance labels are based on lexical matching, the corpus is limited to ChemRxiv papers, and the study evaluates retrieval quality rather than full end-to-end RAG answer generation. Strong retrieval may improve the availability of evidence for downstream generation, but it does not by itself guarantee factuality, citation accuracy, or chemical reasoning quality. The chunking grid was also constrained by the evaluated token range, and the results should not be generalized to all possible semantic, adaptive, learned, or long-context chunking methods. Future work should include manual qrel validation, uncertainty analysis, hybrid sparse-dense retrieval, reranking, larger context windows, and full RAG evaluation with generated-answer quality and citation faithfulness.

\section{Conclusion}

This work presented a systematic benchmark study of chunking-strategy and embedding-model trade-offs for chemistry evidence retrieval in retrieval-augmented generation. By combining external chemistry-oriented retrieval benchmarks with ChemQuests-derived retrieval tasks, the study provides practical evidence on how embedding choice, chunking strategy, chunk size, and overlap shape retrieval quality and computational efficiency.

The results identify retrieval-tuned embedding models as strong baselines for chemistry evidence retrieval, with embedding choice associated with the largest observed performance differences. Within the evaluated grid, large chunks with low or zero overlap provided a practical starting point, and fixed-token, recursive-token, and hierarchical-section chunking offered effective and straightforward options for chemistry-aware retrieval pipelines.

The study also contributes a reusable ChemQuests-derived retrieval benchmark family and uses \(\mathrm{Geom@10}\) as a compact screening criterion for selecting strong candidate embedding models before the larger chunking evaluation. These results provide empirical guidance for selecting retrieval configurations for chemistry-focused RAG systems.

Future work can build on this foundation through manual qrel validation, statistical uncertainty analysis, broader chunk-size ranges, sanity checks with additional non-shortlisted models, and full end-to-end RAG evaluation.


\section*{Acknowledgements}
The authors gratefully acknowledge support from the BMFTR within the funding program Photonics Research Germany through projects 13N15466 (LPI-BT1-FSU), 13N15710 (LPI-BT3-FSU), and 13N15715 (LPI-BT4-FSU). The authors also thank their colleagues at the Leibniz Institute of Photonic Technology and Friedrich Schiller University Jena for valuable discussions and support, and acknowledge the open-source community whose tools and resources made this work possible.

\section*{Funding}
This work was supported by the BMFTR funding program Photonics Research Germany (13N15466 (LPI-BT1-FSU), 13N15710 (LPI-BT3-FSU), 13N15715 (LPI-BT4-FSU)) and is integrated into the Leibniz Centre for Photonics in Infection Research (LPI). The LPI, initiated by Leibniz IPHT, Leibniz-HKI, Friedrich Schiller University Jena, and Jena University Hospital, is part of the BMFTR national roadmap for research infrastructures.

\section*{Competing interests}
The authors declare no competing interests.

\section*{Ethics approval and consent to participate}
Not applicable.

\section*{Consent for publication}
Not applicable.

\section*{Availability of data and materials}
The benchmark data and materials generated and analyzed during this study are publicly available in the Bocklitz Lab GitHub repository at \url{https://github.com/Bocklitz-Lab/chunk_twice_embed_once}.

\section*{Code availability}
The source code for the retrieval pipeline, chunking framework, and evaluation scripts used in this work is publicly available at the Bocklitz Lab GitHub repository, \url{https://github.com/Bocklitz-Lab/chunk_twice_embed_once}.

\section*{Authors\textquotesingle{} contributions}
Conceptualization, M.A. and T.B.; methodology, M.A. and T.B.; validation, M.A. and T.B.; formal analysis, M.A.; investigation, M.A.; resources, M.A.; data curation, M.A.; writing---original draft preparation, M.A.; writing---review and editing, M.A. and T.B.; visualization, M.A.; supervision, T.B.; project administration, T.B.; funding acquisition, T.B. All authors have read and approved the final manuscript.

\clearpage
\setcounter{section}{0}
\setcounter{figure}{0}
\setcounter{table}{0}
\setcounter{equation}{0}
\setcounter{algorithm}{0}
\renewcommand{\thefigure}{S\arabic{figure}}
\renewcommand{\thetable}{S\arabic{table}}
\renewcommand{\theequation}{S\arabic{equation}}
\renewcommand{\thealgorithm}{S\arabic{algorithm}}
\renewcommand{\theHsection}{supp.\arabic{section}}
\renewcommand{\theHfigure}{supp.S\arabic{figure}}
\renewcommand{\theHtable}{supp.S\arabic{table}}
\renewcommand{\theHequation}{supp.S\arabic{equation}}
\providecommand{\theHalgorithm}{}
\renewcommand{\theHalgorithm}{supp.S\arabic{algorithm}}

\section{Supplementary Materials}
\label{sec:supplementary}

\subsection{Related Work on Chunking Strategies}
\label{sec:supp_related_chunking}

Linear text segmentation, or topic segmentation, identifies boundaries at which the topic changes. It is relevant to retrieval because these boundaries determine the text units available for indexing and ranking~\cite{ghinassi2024recent}. Early unsupervised methods used lexical cohesion, divisive clustering, dynamic programming, or probabilistic cues. Examples include TextTiling~\cite{hearst1994multi}, C99~\cite{choi2000linear}, U00~\cite{utiyama2001statistical}, and BayesSeg~\cite{barzilay2008modeling}. Topic-model-based methods instead infer latent topic distributions to identify segment boundaries~\cite{misra2011text, riedl2012text}.

Embedding-based methods use similarities between local text representations to identify changes in semantic content. GraphSeg~\cite{glavavs2016unsupervised} and BERT-based methods~\cite{solbiati2021unsupervised} are examples of this approach.

Supervised segmentation systems formulate boundary detection as sequence labeling or classification. TextSeg~\cite{koshorek2018text}, Cross-Segment BERT~\cite{lukasik2020text}, and Longformer-based TSSP and CSSL~\cite{yu2023improving} use contextual representations to model transitions between text units. Multi-task approaches combine segmentation with objectives such as topic classification or coherence modeling~\cite{lo2021transformer, barrow2020joint, somasundaran2020two}; Tipster integrates neural topic modeling with segmentation~\cite{gong2022tipster}. The extent to which these supervised methods transfer to chemistry documents remains unclear.

Large language models have also been evaluated for zero-shot segmentation through prompting. Fan and Jiang~\cite{fan2023uncovering} and Jiang et al.~\cite{jiang2023superdialseg} report competitive performance on dialogue-segmentation tasks. These results do not establish equivalent performance for chemistry articles, which differ in terminology, notation, and document structure.

In retrieval-augmented generation (RAG), segmentation defines the evidence units indexed by the retriever. Separating related evidence across chunk boundaries can reduce the completeness of retrieved context and may affect the evidence supplied to a downstream generator. The present study evaluates retrieval quality only; generated-answer quality was not measured.

\subsection{Related Work on Embedding Models}
\label{sec:supp_related_embedding_models}

Text representations used for retrieval include sparse count-based vectors, static word embeddings, contextual token representations, and sentence- or passage-level embeddings~\cite{cao2024recent}. Bag-of-Words and Term Frequency--Inverse Document Frequency (TF-IDF) represent text using sparse term counts or weights~\cite{petukhova2024text}. Word2Vec~\cite{mikolov2013efficient}, GloVe~\cite{pennington2014glove}, and FastText~\cite{bojanowski2017enriching} provide dense but context-invariant word representations. Contextual models such as ELMo~\cite{peters2018deep}, BERT~\cite{devlin2019bert}, and GPT~\cite{radford2018improving} instead condition token representations on surrounding text.

Dense retrieval requires query and document representations that can be compared in a shared vector space. Sentence-BERT (SBERT) adapted BERT with Siamese and triplet architectures for sentence-level similarity~\cite{reimers2019sentence}. The evaluated model set includes Sentence-Transformers variants based on BERT, MiniLM, and MPNet, including bert-base-nli-mean-tokens\footnote{\url{https://huggingface.co/sentence-transformers/bert-base-nli-mean-tokens}}~\cite{devlin2019bert}, all-MiniLM-L6-v2\footnote{\url{https://huggingface.co/sentence-transformers/all-MiniLM-L6-v2}}~\cite{wang2020minilm}, all-mpnet-base-v2\footnote{\url{https://huggingface.co/sentence-transformers/all-mpnet-base-v2}}~\cite{song2020mpnet}, and multi-qa-mpnet-base-dot-v1\footnote{\url{https://huggingface.co/sentence-transformers/multi-qa-mpnet-base-dot-v1}}~\cite{song2020mpnet}.

Retrieval-oriented embedding families include E5~\cite{wang2022text}, GTE~\cite{li2023towards}, and BGE~\cite{xiao2024c}. E5 uses large-scale contrastive training data, and BGE uses retrieval-oriented training that includes negative examples and question-answer data. The multitask bge-m3 model\footnote{\url{https://huggingface.co/BAAI/bge-m3}} supports several embedding and retrieval settings~\cite{xu2024m3embedding}. These models are directly relevant to the present study because the experiments compare their retrieval behavior with that of general-purpose and domain-pretrained encoders.

The evaluated domain-pretrained models include SciBERT~\cite{beltagy2019scibert}, PubMedBERT~\cite{gu2021domain}, BiomedBERT~\cite{biomedbert}, ChemicalBERT~\cite{recobo2023chembert}, ChemBERTa~\cite{chemberta}, and MatSciBERT~\cite{gupta_matscibert_2022}. Their pretraining corpora contain scientific, biomedical, chemical, or materials-science text, but domain pretraining does not itself establish retrieval performance. The experiments therefore compare these models directly with retrieval-tuned encoders.

The model set also includes Nomic Embed~\cite{nomic2024embed} and multilingual E5 and BGE variants~\cite{wang2024multilingual, xu2024m3embedding}. Including these families broadens the Phase 1 comparison across general-purpose, retrieval-tuned, domain-pretrained, instruction-tuned, and multilingual encoders.

\subsection{External Chemistry Retrieval Tasks}
\label{sec:supp_external_chemistry_retrieval_tasks}

To assess model transfer across two external chemistry retrieval settings, we benchmarked all models on two chemistry-focused tasks from the Massive Text Embedding Benchmark (MTEB). Both datasets are publicly available through Hugging Face.

\textbf{ChemHotpotQARetrieval}\footnote{\url{https://huggingface.co/datasets/BASF-AI/ChemHotpotQARetrieval}} is a multi-hop retrieval task derived from the HotpotQA\footnote{\url{https://huggingface.co/datasets/mteb/hotpotqa}} dataset, which consists of question-answer pairs sourced from English Wikipedia. For this chemistry-specific subset, questions were filtered by identifying entries within the chemistry category and traversing linked articles up to three levels deep. The task evaluates retrieval of evidence distributed across multiple documents for chemistry questions.

\textbf{ChemNQRetrieval}\footnote{\url{https://huggingface.co/datasets/BASF-AI/ChemNQRetrieval}} is based on the Chemistry Natural Questions subset of the Natural Questions dataset\footnote{\url{https://huggingface.co/datasets/mteb/nq}}, which contains real user queries issued to Google and grounded in Wikipedia content. Similar to ChemHotpotQA, chemistry-related questions were extracted through a structured traversal of Wikipedia's chemistry category and its linked pages. This dataset reflects real-world information needs in an open-domain chemistry context.

The two tasks provide complementary screening settings: multi-hop document retrieval and retrieval for open-domain chemistry questions. They are used here to compare model transfer before the ChemQuests evaluation.

\subsection{Detailed Chunking Method Definitions}
\label{sec:supp_chunking_methods}

\begin{figure}[!htbp]
\centering
\includegraphics[width=\textwidth,height=0.72\textheight,keepaspectratio]{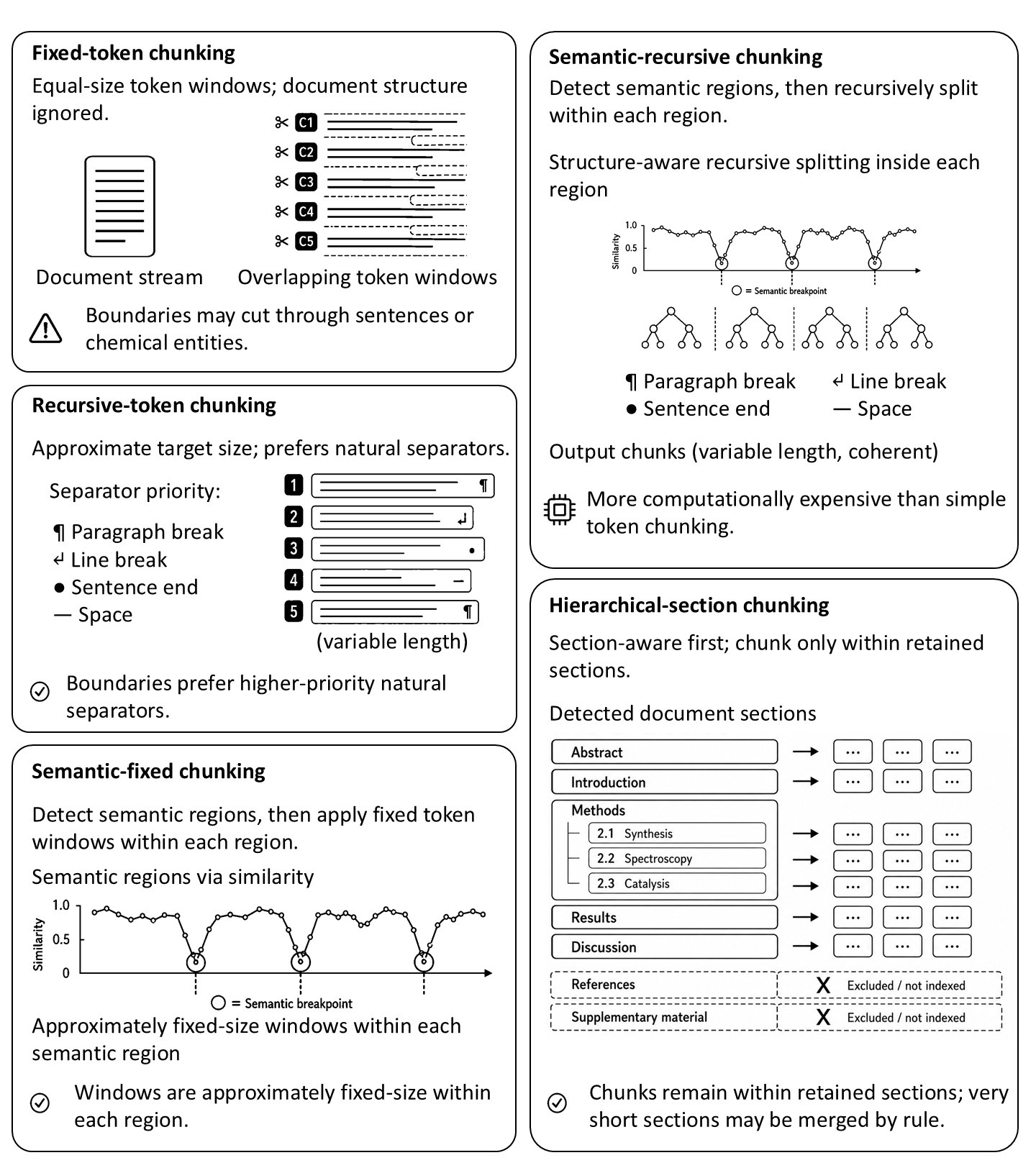}
\caption{Overview of the five chunking strategies evaluated in this study, ranging from uniform token windows to semantic and section-constrained segmentation.}
\label{fig:supp_chunking_methods}
\end{figure}

Let a document be represented as an ordered token sequence
\[
D = (t_1, t_2, \dots, t_n).
\]
A chunking method defines a collection of contiguous spans extracted from $D$
\[
\mathcal{C} = \{c_1, c_2, \dots, c_m\}, \qquad
c_i = (t_{a_i}, \dots, t_{b_i}),
\]
where each chunk $c_i$ satisfies a length constraint and adjacent chunks may overlap. In retrieval-augmented generation, the chunking function is important because it determines the atomic units indexed by the retriever. The retrieval objective is therefore affected not only by the embedding function $f(\cdot)$, but also by the segmentation operator $S(\cdot)$ that maps a document to chunks.
\[
D \xrightarrow{S} \mathcal{C} \xrightarrow{f} \{f(c_1), \dots, f(c_m)\}.
\]
The chunking strategies considered here differ in how they balance evidence continuity against the granularity of the indexed units.

In this study, we evaluate five chunking strategies that represent distinct segmentation principles, namely fixed-token chunking, recursive-token chunking, semantic-fixed chunking, semantic-recursive chunking, and hierarchical-section chunking. These methods differ in the criterion used to place boundaries, namely regular token intervals, syntactic or discourse cues, semantic change, or document structure. Here, token length is defined using the Hugging Face tokenizer associated with the embedding model in the corresponding experimental condition. Consequently, a nominal chunk size such as 256 tokens is model-tokenizer-dependent and may yield slightly different chunk boundaries across embedding models.

\paragraph{Fixed-token chunking.}
The fixed-token strategy partitions the document into chunks of approximately constant length $L$, optionally with overlap $o$. Formally, chunk boundaries are placed at regular intervals, so that the $i$th chunk begins near position
\[
a_i = 1 + (i-1)(L-o),
\]
and spans roughly $L$ tokens. This strategy is simple, computationally efficient, and yields uniform retrieval units. However, its boundaries are independent of linguistic or scientific structure. As a result, it may fragment sentences, split definitions from supporting context, or separate experimental conditions from their outcomes.

\paragraph{Recursive-token chunking.}
Recursive-token chunking introduces boundary sensitivity by preferring segmentation at increasingly fine textual units while still enforcing a target token budget $L$. Let the ordered separator hierarchy be
\[
\Sigma = (\text{\textbackslash n\textbackslash n},\ \text{\textbackslash n},\ \text{.},\ \text{?},\ \text{!},\ \text{space},\ \text{""}).
\]
The splitter selects the earliest separator in $\Sigma$ for which the resulting span satisfies the target budget $L$; if no earlier separator is feasible, the procedure falls back to progressively finer separators. Delimiters are retained in the resulting chunks, while surrounding whitespace is stripped. Conceptually, this method seeks a segmentation that still respects a target chunk budget $L$ but minimizes disruption of natural discourse boundaries. It can be viewed as an optimization over candidate splits.
\[
S^* = \arg\min_{S \in \mathcal{S}_L} \sum_{b \in S} \phi(b),
\]
where $\mathcal{S}_L$ denotes the set of segmentations satisfying the length constraint and $\phi(b)$ is a penalty for placing a boundary at an undesirable location. Boundaries at paragraph or sentence breaks have lower penalty than boundaries inside a sentence or phrase.

\paragraph{Semantic chunking.}
Semantic chunking aims to place boundaries where topical or conceptual continuity decreases. Let $u_1, u_2, \dots, u_k$ denote a sequence of local text units, and let $e(u_j)$ be their vector representations. A semantic segmentation method identifies candidate breakpoints by analyzing the similarity between neighboring units, for example through
\[
s_j = \cos\!\bigl(e(u_j), e(u_{j+1})\bigr).
\]
Low values of $s_j$ can indicate a possible semantic transition and therefore a plausible chunk boundary. More generally, semantic chunking attempts to maximize within-chunk coherence while separating distinct topics across chunks. This can be expressed abstractly as favoring segmentations for which
\[
\sum_{c \in \mathcal{C}} \mathrm{coh}(c)
\]
is large, where $\mathrm{coh}(c)$ denotes the semantic homogeneity of chunk $c$.

In this study, semantic chunking is formulated as a two-stage procedure. Candidate semantic breakpoints are first identified by an embedding-based semantic splitter using percentile thresholding with threshold level $\tau = 95$, buffer parameter $b = 1$, and minimum segment-size parameter $m_{\min} = 128$. A size-constrained chunking operator is then applied to these semantically delimited regions to produce retrieval-ready chunks. We evaluate two semantic chunking strategies. In \textit{semantic-fixed}, semantic regions are detected first and the size constraint is then enforced using fixed token windows. In \textit{semantic-recursive}, semantic regions are likewise detected first, but the size constraint is enforced using recursive structural splitting rather than uniform token windows. These two chunking strategies therefore differ in the post hoc size-enforcement stage applied after semantic region detection.

\paragraph{Hierarchical-section chunking.}
Scientific articles are organized hierarchically into sections and subsections that reflect discourse function. Hierarchical-section chunking exploits this structure by first partitioning the document according to its discourse hierarchy and then segmenting within those units. If the document is decomposed into sections
\[
D = (D_1, D_2, \dots, D_r),
\]
where $(D_1,\dots,D_r)$ denotes an ordered partition of the document into sections, then chunking is applied within each $D_\ell$ rather than across section boundaries. This constrains chunks to remain locally consistent with the rhetorical and scientific function of the surrounding text.

In this study, section boundaries are detected by a rule-based procedure combining LaTeX section commands, numbered headings, chemistry-specific heading patterns, generic heading heuristics, and custom regular-expression rules. Sections shorter than the minimum section-length threshold $s_{\min} = 128$ are merged with adjacent sections, with forward merging preferred when feasible. Once end-matter headings are detected, subsequent sections corresponding to references, bibliography, acknowledgments, author information, supplementary or supporting information, data availability, abbreviations, nomenclature, glossary material, and related back matter are excluded from indexing.

\paragraph{Comparison of the chunking principles.}
The five evaluated chunking strategies can be understood as imposing different priors on where informative boundaries occur. Fixed-token chunking assumes no privileged boundary locations and prioritizes uniformity, although the realized token spans remain conditional on the model-specific tokenizer. Recursive-token chunking assumes that linguistic structure provides useful cues for preserving local coherence. The evaluated semantic chunking implementations assume that topic shifts are the most important signals for segmentation, with size control applied after semantic region detection. Hierarchical-section chunking assumes that document-level discourse structure is a reliable guide to chunk boundaries. These assumptions lead to different trade-offs between granularity, coherence, and retrieval specificity.

More generally, chunking can be viewed as balancing two competing objectives. Smaller chunks are often expected to improve precision because they localize evidence more sharply, whereas larger chunks may improve recall and contextual completeness because they preserve more surrounding information. Overlap can partially mediate this trade-off by reducing the probability that relevant evidence is split across chunk boundaries.

\subsection{Retrieval Metric Definitions}
\label{sec:supp_retrieval_metrics}

We evaluate retrieval performance using the standard metrics employed in the \textsc{MTEB} retrieval pipeline. For each query $q \in Q$, let $G_q$ denote the set of relevant chunks, let $\pi_q(i)$ denote the chunk ranked at position $i$, and define the binary relevance indicator
\[
r_q(i)=
\begin{cases}
1, & \text{if } \pi_q(i)\in G_q,\\
0, & \text{otherwise.}
\end{cases}
\]
Following the \textsc{MTEB}/Sentence-Transformers retrieval evaluator, all metrics are computed per query and then macro-averaged over the query set $Q$. Any chunk not contained in $G_q$ is treated as non-relevant, consistent with the standard \textsc{MTEB} retrieval evaluation protocol.

\paragraph{Normalized Discounted Cumulative Gain (\texorpdfstring{\(\mathrm{nDCG@}k\)}{nDCG@k}).}
\[
\mathrm{DCG@}k(q)=\sum_{i=1}^{k}\frac{r_q(i)}{\log_2(i+1)}.
\]
\[
\mathrm{IDCG@}k(q)=\sum_{i=1}^{\min(k,|G_q|)}\frac{1}{\log_2(i+1)}.
\]
\[
\mathrm{nDCG@}k(q)=\frac{\mathrm{DCG@}k(q)}{\mathrm{IDCG@}k(q)},
\qquad
\mathrm{nDCG@}k=\frac{1}{|Q|}\sum_{q\in Q} \mathrm{nDCG@}k(q).
\]

\paragraph{Mean Average Precision (\texorpdfstring{\(\mathrm{MAP@}k\)}{MAP@k}).}
\[
\mathrm{P@}i(q)=\frac{1}{i}\sum_{j=1}^{i} r_q(j).
\]
\[
\mathrm{AP@}k(q)=\frac{1}{\min(k,|G_q|)}\sum_{i=1}^{k} \mathrm{P@}i(q)\,r_q(i),
\]
\[
\mathrm{MAP@}k=\frac{1}{|Q|}\sum_{q\in Q} \mathrm{AP@}k(q).
\]

\paragraph{\texorpdfstring{\(\mathrm{Precision@}k\)}{Precision@k}.}
\[
\mathrm{Precision@}k(q)=\frac{1}{k}\sum_{i=1}^{k} r_q(i),
\qquad
\mathrm{Precision@}k=\frac{1}{|Q|}\sum_{q\in Q} \mathrm{Precision@}k(q).
\]

\paragraph{\texorpdfstring{\(\mathrm{Recall@}k\)}{Recall@k}.}
\[
\mathrm{Recall@}k(q)=\frac{1}{|G_q|}\sum_{i=1}^{k} r_q(i),
\qquad
\mathrm{Recall@}k=\frac{1}{|Q|}\sum_{q\in Q} \mathrm{Recall@}k(q).
\]

\paragraph{Mean Reciprocal Rank (\texorpdfstring{\(\mathrm{MRR@}k\)}{MRR@k}).}
Let
\[
rank_q^\star=\min\{\, i \leq k : r_q(i)=1 \,\},
\]
with $\mathrm{RR@}k(q)=0$ if no relevant chunk appears in the top-$k$ list. The reciprocal rank is
\[
\mathrm{RR@}k(q)=
\begin{cases}
\frac{1}{rank_q^\star}, & \text{if a relevant chunk appears in the top } k,\\
0, & \text{otherwise,}
\end{cases}
\]
and the mean reciprocal rank is
\[
\mathrm{MRR@}k=\frac{1}{|Q|}\sum_{q\in Q} \mathrm{RR@}k(q).
\]

\paragraph{Composite Metric (\texorpdfstring{\(\mathrm{Geom@10}\)}{Geom@10}).}
\[
\mathrm{Geom@10}=\sqrt{\mathrm{nDCG@10}\cdot \mathrm{Recall@10}}.
\label{eq:supp_geom10}
\]

\paragraph{Evaluation Time.}
In addition to effectiveness metrics, we report evaluation time as an efficiency measure. Let $t_{\mathrm{start}}$ and $t_{\mathrm{end}}$ denote the start and end timestamps of the evaluation procedure for one configuration. We define
\[
T_{\mathrm{eval}} = t_{\mathrm{end}} - t_{\mathrm{start}}.
\]

\paragraph{Remark on \texorpdfstring{\(\mathrm{Accuracy@}k\)}{Accuracy@k}.}
Although the underlying evaluator can also compute $\mathrm{Accuracy@}k$, we do not emphasize it in the main analysis. In retrieval settings, $\mathrm{Accuracy@}k$ only measures whether at least one relevant chunk appears in the top-$k$ list and does not describe ranking quality, relevance concentration, or coverage. Downstream generation was not evaluated in this study.

\subsection{Evaluated Embedding Models}
\label{sec:supp_evaluated_embedding_models}

The evaluated embedding-model set included Sentence-Transformers, Microsoft biomedical and general encoders, E5, BGE, Nomic, Contriever, ChemBERTa, BERT, MatSciBERT, Instructor, SciBERT, SPECTER, and ChemBERT variants~\cite{reimers2021allminilm,reimers2019sentence,reimers2021allminilm12v2,reimers2021allmpnet,reimers2021gtrt5base,reimers2021gtrt5large,reimers2021gtrt5xl,reimers2019bertbase,reimers2021multiqa,reimers2021paraphrasempnet,gu2021domain,microsoft2021biomedbert,microsoft2021biomedbertabstract,wang2020minilm,microsoft2020minilmh384,song2020mpnet,microsoft2020mpnet,wang2022text,wang2022e5small,wang2022e5smallv2,wang2022e5base,wang2022e5basev2,wang2022e5large,wang2022e5largev2,wang2022multilinguale5base,wang2022multilinguale5large,wang2022multilinguale5small,bge_embedding,baai2023bgesmall,baai2023bgesmallv15,baai2023bgebase,baai2023bgebasev15,baai2023bgelarge,baai2023bgelargev15,xu2024m3embedding,baai2023bgem3,nomicai2023embedv1,nomic2024embed,nomicai2024embedv15,nomicai2023bert2048,shazeer2020glu,izacard2021unsupervised,izacard2022contriever,izacard2022contrievermsmarco,chemberta,deepchem2023chemberta77m,matscibert2021,gupta_matscibert_2022,hkunlp2022instructorxl,su2022one,beltagy2019scibert,cohan2020specter,allenai2020specter,recobo2023chembert}.

Table~\ref{tab:decomposed_model_name_map} provides the corresponding full model identifiers, library families, size or version information, and the abbreviated names used in figures and compact summaries throughout the manuscript.

\begin{scriptsize}
    \begin{longtable}{@{}p{0.15\textwidth}p{0.25\textwidth}p{0.11\textwidth}p{0.09\textwidth}p{0.25\textwidth}@{}}
    \caption{Decomposed Model Identifiers and Their Abbreviated Short Names. This table lists the full Hugging Face model identifiers along with their corresponding libraries, size or architecture details, version numbers (if applicable), and standardized short names used throughout the paper. These short names are referenced in various sections to simplify model comparison and discussion.}
    \label{tab:decomposed_model_name_map} \\
    \toprule
    Library & Model & Size/Layer & Version & Short Name \\
    \midrule
    \endfirsthead

    \multicolumn{5}{c}{{\bfseries \tablename\ \thetable{} -- continued from previous page}} \\
    \toprule
    Library & Model & Size/Layer & Version & Short Name \\
    \midrule
    \endhead

    \midrule \multicolumn{5}{r}{{Continued on next page}} \\
    \endfoot

    \bottomrule
    \endlastfoot
    sentence-transformers & all-MiniLM & L6 & v2 & Sent\_MiniLM6\_v2~\cite{reimers2021allminilm, reimers2019sentence} \\
    sentence-transformers & all-MiniLM & L12 & v2 & Sent\_MiniLM12\_v2~\cite{reimers2021allminilm12v2, reimers2019sentence} \\
    sentence-transformers & all-mpnet-base & -- & v2 & Sent\_allMPNetB\_v2~\cite{reimers2021allmpnet, reimers2019sentence} \\
    sentence-transformers & gtr-t5 & base & -- & Sent\_gtrT5\_B~\cite{reimers2021gtrt5base, reimers2019sentence} \\
    sentence-transformers & gtr-t5 & large & -- & Sent\_gtrT5\_L~\cite{reimers2021gtrt5large, reimers2019sentence} \\
    sentence-transformers & gtr-t5 & xl & -- & Sent\_gtrT5\_XL~\cite{reimers2021gtrt5xl, reimers2019sentence} \\
    sentence-transformers & bert-base-nli-mean-tokens & -- & -- & Sent\_BERTnli~\cite{reimers2019bertbase, reimers2019sentence} \\
    sentence-transformers & multi-qa-mpnet-base-dot & v1 & -- & Sent\_MQA\_MPNetB~\cite{reimers2021multiqa, reimers2019sentence} \\
    sentence-transformers & paraphrase-multilingual-mpnet-base & -- & v2 & Sent\_paraMPNetB\_v2~\cite{reimers2021paraphrasempnet, reimers2019sentence} \\
    microsoft & BiomedNLP-BiomedBERT-base-uncased-abstract-fulltext & -- & -- & MS\_BiomedBERT\_af~\cite{gu2021domain, microsoft2021biomedbert} \\
    microsoft & BiomedNLP-BiomedBERT-base-uncased-abstract & -- & -- & MS\_BiomedBERT\_a~\cite{gu2021domain, microsoft2021biomedbertabstract} \\
    microsoft & MiniLM & L12-H384 & -- & MS\_MiniLM384~\cite{wang2020minilm, microsoft2020minilmh384} \\
    microsoft & mpnet & base & -- & MS\_MPNetB~\cite{song2020mpnet, microsoft2020mpnet} \\
    intfloat & e5 & small & v1 & Intf\_e5S~\cite{wang2022text, wang2022e5small} \\
    intfloat & e5 & small & v2 & Intf\_e5S\_v2~\cite{wang2022text, wang2022e5smallv2} \\
    intfloat & e5 & base & v1 & Intf\_e5B~\cite{wang2022text, wang2022e5base} \\
    intfloat & e5 & base & v2 & Intf\_e5B\_v2~\cite{wang2022text, wang2022e5basev2} \\
    intfloat & e5 & large & v1 & Intf\_e5L~\cite{wang2022text, wang2022e5large} \\
    intfloat & e5 & large & v2 & Intf\_e5L\_v2~\cite{wang2022text, wang2022e5largev2} \\
    intfloat & multilingual-e5 & base & -- & Intf\_mE5B~\cite{wang2022text, wang2022multilinguale5base} \\
    intfloat & multilingual-e5 & large & -- & Intf\_mE5L~\cite{wang2022text, wang2022multilinguale5large} \\
    intfloat & multilingual-e5 & small & -- & Intf\_mE5S~\cite{wang2022text, wang2022multilinguale5small} \\
    BAAI & bge & small-en & -- & BAAI\_bgeS~\cite{bge_embedding, baai2023bgesmall} \\
    BAAI & bge & small-en & v1.5 & BAAI\_bgeS\_v1.5~\cite{bge_embedding, baai2023bgesmallv15} \\
    BAAI & bge & base-en & -- & BAAI\_bgeB~\cite{bge_embedding, baai2023bgebase} \\
    BAAI & bge & base-en & v1.5 & BAAI\_bgeB\_v1.5~\cite{bge_embedding, baai2023bgebasev15} \\
    BAAI & bge & large-en & -- & BAAI\_bgeL~\cite{bge_embedding, baai2023bgelarge} \\
    BAAI & bge & large-en & v1.5 & BAAI\_bgeL\_v1.5~\cite{bge_embedding, baai2023bgelargev15} \\
    BAAI & bge & m3 & -- & BAAI\_bgeM3~\cite{xu2024m3embedding, baai2023bgem3} \\
    nomic-ai & nomic-embed-text & -- & v1 & Nomic\_text\_v1~\cite{nomicai2023embedv1, nomic2024embed} \\
    nomic-ai & nomic-embed-text & -- & v1.5 & Nomic\_text\_v1.5~\cite{nomicai2024embedv15, nomic2024embed} \\
    nomic-ai & nomic-embed-text-v2-moe & -- & v2 & Nomic\_text\_v2~\cite{nomic2024embed} \\
    nomic-ai & nomic-bert & 2048 & -- & Nomic\_BERT2048~\cite{nomicai2023bert2048, shazeer2020glu} \\
    facebook & contriever & -- & -- & FB\_contriever~\cite{izacard2021unsupervised, izacard2022contriever} \\
    facebook & contriever-msmarco & -- & -- & FB\_contrieverMS~\cite{izacard2021unsupervised, izacard2022contrievermsmarco} \\
    DeepChem & ChemBERTa & 77M & MTR & DC\_ChemBERTa~\cite{chemberta, deepchem2023chemberta77m} \\
    m3rg-iitd & matscibert & -- & -- & IITD\_MatSci~\cite{matscibert2021, gupta_matscibert_2022} \\
    hkunlp & instructor-xl & xl & -- & HKU\_InstructXL~\cite{hkunlp2022instructorxl, su2022one} \\
    allenai & scibert-scivocab-uncased & base & -- & AI\_SciBERT~\cite{beltagy2019scibert} \\
    allenai & specter & -- & -- & AI\_Specter~\cite{cohan2020specter, allenai2020specter} \\
    recobo & chemical-bert-uncased & -- & -- & Rec\_ChemBERT~\cite{recobo2023chembert} \\
    \end{longtable}
\end{scriptsize}
\subsection{Complete Phase 1 Embedding-Model Screening Results}
\label{sec:supp_phase1_full_screening}

\begin{figure}[!htbp]
    \centering
    \makebox[\textwidth][c]{%
        \includegraphics[width=1.2\textwidth]{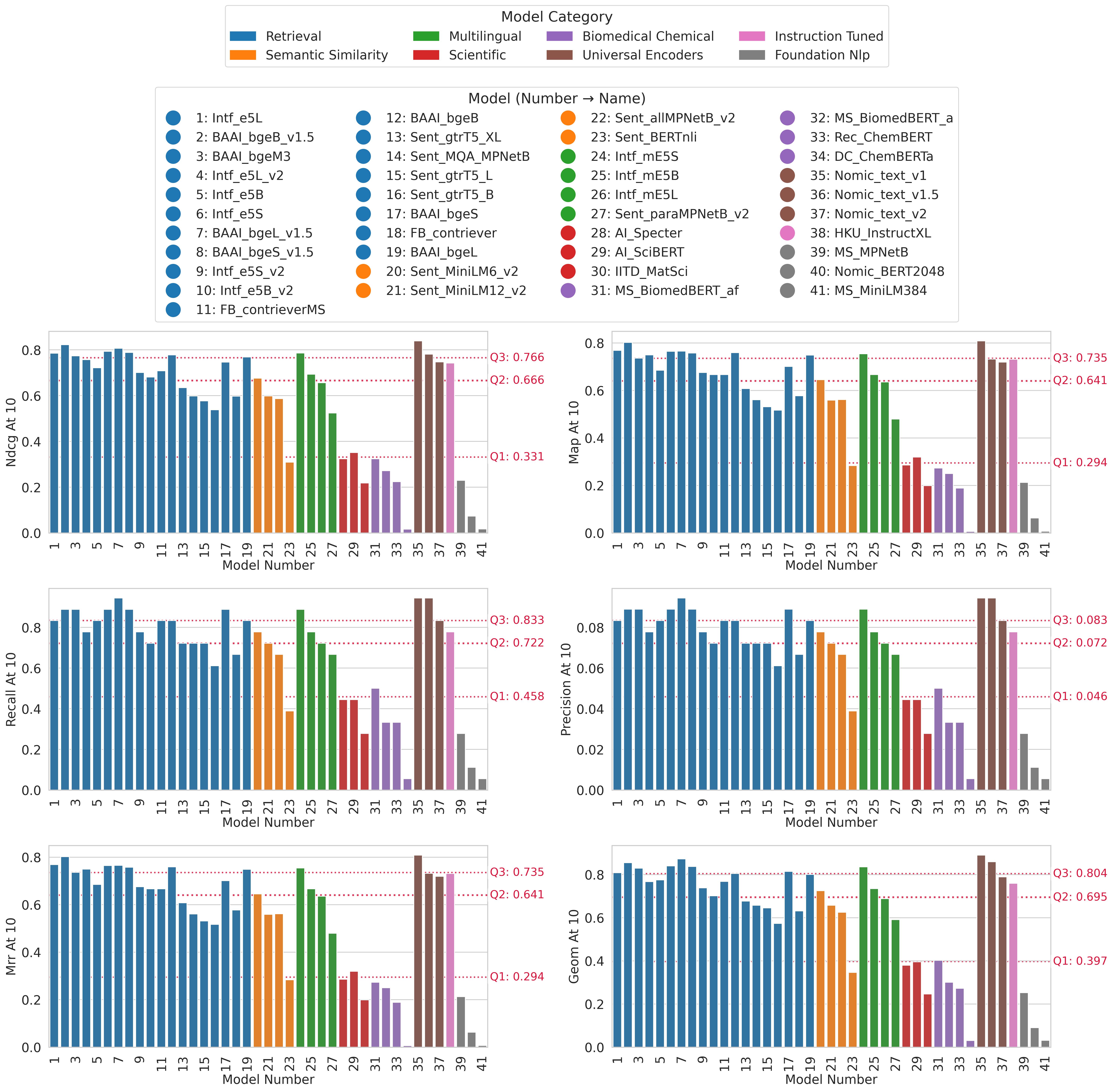}%
    }
    \caption{Phase 1 screening results on {ChemHotpotQARetrieval}. Bars report six retrieval metrics for all 41 evaluated embedding models, namely \(\mathrm{nDCG@10}\), \(\mathrm{MAP@10}\) (mean average precision at rank 10), \(\mathrm{Recall@10}\), \(\mathrm{Precision@10}\) (precision at rank 10), \(\mathrm{MRR@10}\) (mean reciprocal rank at rank 10), and the composite screening score \(\mathrm{Geom@10}\) defined in the main text. Models are indexed on the x-axis and mapped to names in the inset legend; colors indicate model families. Red dotted lines mark the first quartile, median, and third quartile for each metric. Retrieval-oriented and broadly transferable encoders occupy the upper portion of the score distribution, with Nomic, BGE, and E5 variants appearing near the top across multiple metrics.}
    \label{fig:phase1_external_screening_ChemHotPotQARetrieval}
\end{figure}

\begin{figure}[!htbp]
    \centering
    \makebox[\textwidth][c]{%
        \includegraphics[width=1.2\textwidth]{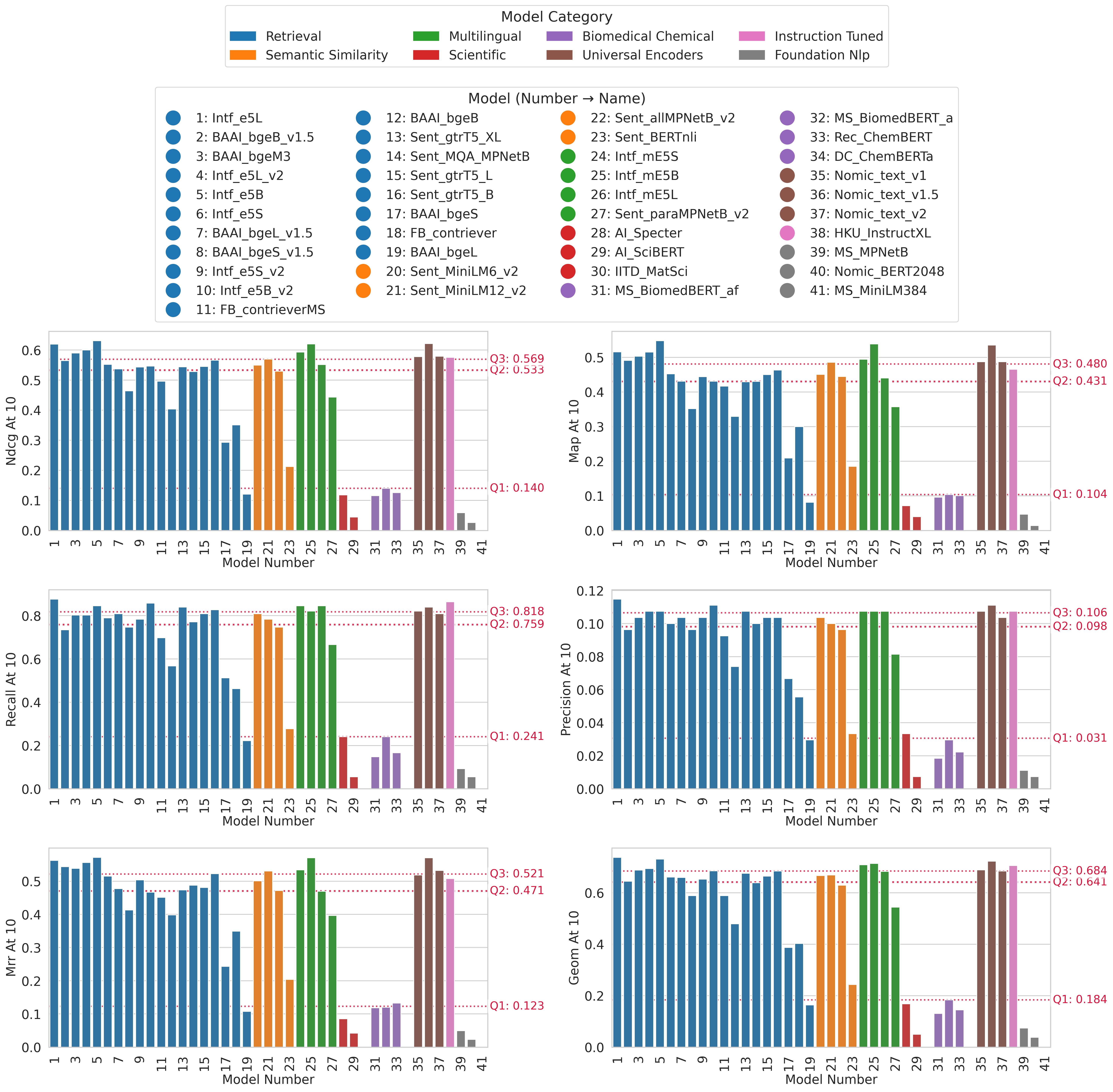}%
    }
    \caption{Phase 1 screening results on {ChemNQRetrieval}. Bars report six retrieval metrics for all 41 evaluated embedding models, namely \(\mathrm{nDCG@10}\), \(\mathrm{MAP@10}\), \(\mathrm{Recall@10}\), \(\mathrm{Precision@10}\), \(\mathrm{MRR@10}\), and the composite screening score \(\mathrm{Geom@10}\) defined in the main text. Models are indexed on the x-axis and mapped to names in the inset legend; colors indicate model families. Red dotted lines mark the first quartile, median, and third quartile for each metric. Compared with {ChemHotpotQARetrieval}, this task more clearly favors E5-family and multilingual E5 encoders, while Nomic and selected BGE models remain competitive in the upper-performance region.}
    \label{fig:phase1_external_screening_ChemNQRetrieval}
\end{figure}

Figures~\ref{fig:phase1_external_screening_ChemHotPotQARetrieval} and~\ref{fig:phase1_external_screening_ChemNQRetrieval} provide the complete Phase 1 screening results for all 41 evaluated embedding models. The corresponding main-text table reports the top-ranked models in compact form.

\begin{table}[htbp]
\centering
\caption{Final Phase 1 shortlist carried forward to Phase 2.}
\label{tab:supp_phase1_shortlist_models}
\begin{tabular}{lp{0.68\textwidth}}
\toprule
Category & Model \\
\midrule
Retrieval & {intfloat/e5-base}~\cite{wang2022text, wang2022e5base} \\
Retrieval & {intfloat/e5-large}~\cite{wang2022text, wang2022e5large} \\
Retrieval & {intfloat/e5-large-v2}~\cite{wang2022text, wang2022e5largev2} \\
Retrieval & {intfloat/e5-small}~\cite{wang2022text, wang2022e5small} \\
Multilingual & {intfloat/multilingual-e5-base}~\cite{wang2022text, wang2024multilingual} \\
Multilingual & {intfloat/multilingual-e5-small}~\cite{wang2022text, wang2024multilingual} \\
Instruction-tuned & {hkunlp/instructor-xl}~\cite{hkunlp2022instructorxl, su2022one} \\
Retrieval & {BAAI/bge-base-en-v1.5}~\cite{bge_embedding, baai2023bgebasev15} \\
Retrieval & {BAAI/bge-large-en-v1.5}~\cite{bge_embedding, baai2023bgelargev15} \\
Retrieval & {BAAI/bge-small-en-v1.5}~\cite{bge_embedding, baai2023bgesmallv15} \\
Retrieval & {BAAI/bge-m3}~\cite{xu2024m3embedding} \\
Universal encoder & {nomic-ai/nomic-embed-text-v1}~\cite{nomicai2023embedv1, nomic2024embed} \\
Universal encoder & {nomic-ai/nomic-embed-text-v1.5}~\cite{nomicai2024embedv15, nomic2024embed} \\
\bottomrule
\end{tabular}
\end{table}

\subsection{Metric Redundancy and Representative Evaluation Criteria}
\label{sec:supp_metric_redundancy}

\begin{figure}[!htbp]
    \centering
    \includegraphics[width=0.8\textwidth]{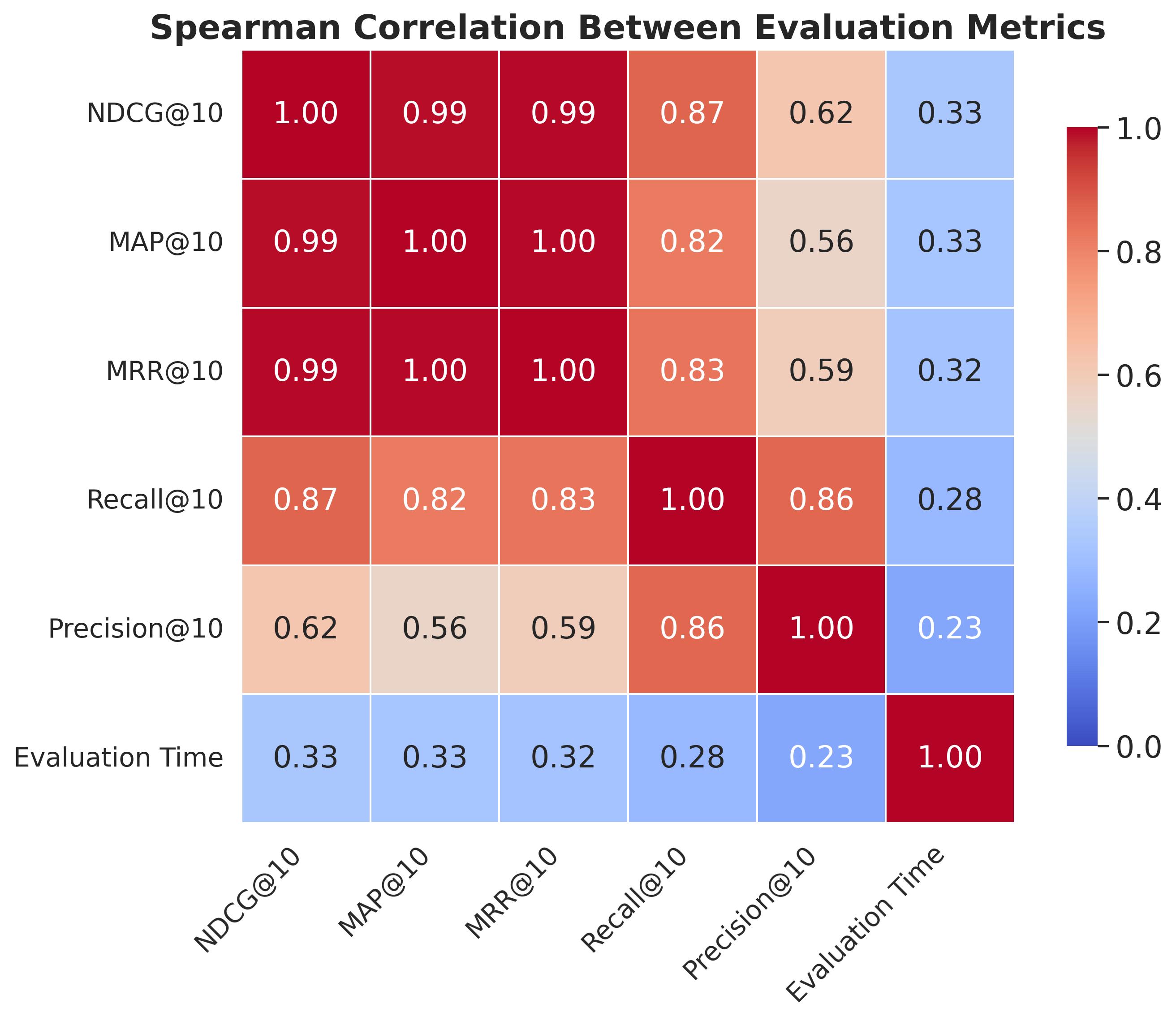}
    \caption{Spearman correlation heatmap for the primary Phase 1 retrieval metrics at cutoff 10 and evaluation time. The rank-sensitive metrics \(\mathrm{nDCG@10}\), \(\mathrm{MAP@10}\), and \(\mathrm{MRR@10}\) form a highly redundant cluster, whereas \(\mathrm{Recall@10}\) and \(\mathrm{Precision@10}\) define a secondary coverage-oriented cluster. Evaluation time shows only weak association with retrieval effectiveness.}
    \label{fig:supp_metric_correlation_phase1}
\end{figure}

\begin{figure}[!htbp]
    \centering
    \includegraphics[width=0.5\textwidth]{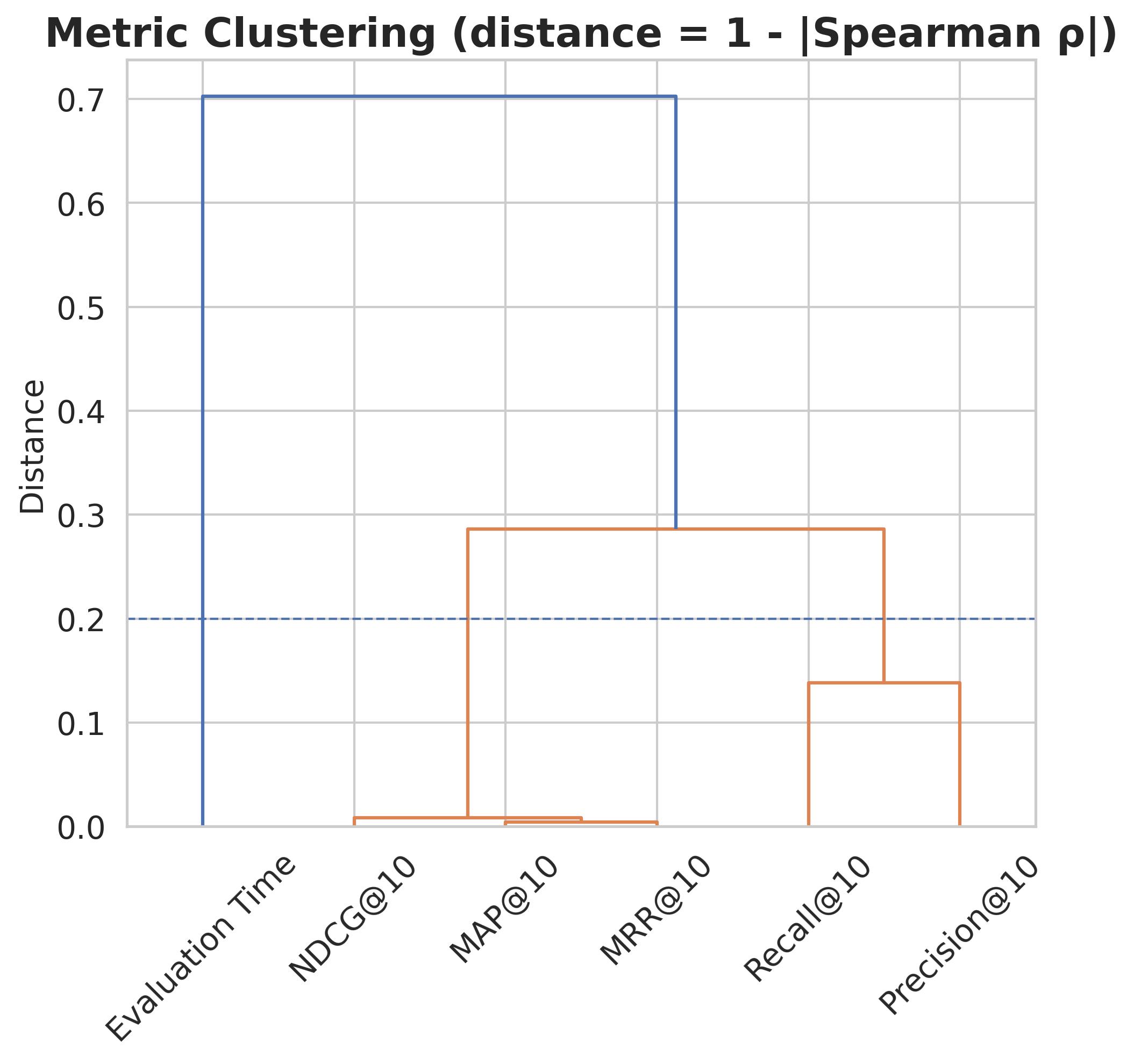}
    \caption{Hierarchical clustering of the primary Phase 1 retrieval metrics using distance \(1-\lvert\rho\rvert\). The clustering confirms the same grouping structure as the correlation heatmap and motivates the use of \(\mathrm{nDCG@10}\) as the representative rank-based metric, \(\mathrm{Recall@10}\) as the representative coverage metric, and evaluation time as the principal efficiency indicator, while \(\mathrm{Geom@10}\) serves as a compact joint summary for Phase 1 model screening.}
    \label{fig:supp_metric_clustering_phase1}
\end{figure}

To interpret the Phase 1 screening results more compactly, we examined redundancy among the primary retrieval metrics at cutoff 10 together with evaluation time. Specifically, we analyzed \(\mathrm{nDCG@10}\), \(\mathrm{MAP@10}\), \(\mathrm{MRR@10}\), \(\mathrm{Recall@10}\), \(\mathrm{Precision@10}\), and total evaluation time over the pooled Phase 1 model-task results. Pairwise relationships were assessed using Spearman rank correlation, and metric groupings were further examined by hierarchical clustering using the distance
\[
d(i,j)=1-\lvert \rho_{ij}\rvert,
\]
where \(\rho_{ij}\) denotes the Spearman correlation between metrics \(i\) and \(j\) (Figures~\ref{fig:supp_metric_correlation_phase1} and~\ref{fig:supp_metric_clustering_phase1}).

The analysis identified a cluster of rank-sensitive metrics. \(\mathrm{nDCG@10}\), \(\mathrm{MAP@10}\), and \(\mathrm{MRR@10}\) were highly correlated in this screening setting. \(\mathrm{Recall@10}\) and \(\mathrm{Precision@10}\) formed a second, coverage-oriented cluster, with \(\mathrm{Recall@10}\) more strongly associated with the rank-based metrics than \(\mathrm{Precision@10}\). Evaluation time was weakly correlated with the effectiveness metrics and was therefore treated as a separate efficiency dimension. Hierarchical clustering produced the same grouping.

These results motivate the use of \(\mathrm{nDCG@10}\) as the representative ranking-quality metric, \(\mathrm{Recall@10}\) as the representative coverage metric, and evaluation time as the principal efficiency indicator. We selected \(\mathrm{nDCG@10}\) from the rank-sensitive cluster because it rewards relevant results near the top of the ranked list while accounting for graded rank position, and it is the default main retrieval score in the \textsc{MTEB} framework. We selected \(\mathrm{Recall@10}\) from the coverage-oriented cluster because the retrieval stage should expose as much answer-relevant evidence as possible to the downstream generator, whereas \(\mathrm{Precision@10}\) is more sensitive to the number of retrieved non-relevant chunks under a fixed cutoff. Under this selection, \(\mathrm{Geom@10}\) serves as a compact joint summary of ranking quality and coverage for Phase 1 screening.

\subsection{Validation of the Composite Metric \texorpdfstring{\(\mathrm{Geom@10}\)}{Geom@10}}
\label{sec:supp_geom10_validation}

Because Phase 1 model selection used \(\mathrm{Geom@10}\), we tested whether this composite preserved the high-performing model region obtained from the full retrieval metric profile.

\begin{figure}[!htbp]
    \centering
    \includegraphics[width=\textwidth,height=0.72\textheight,keepaspectratio]{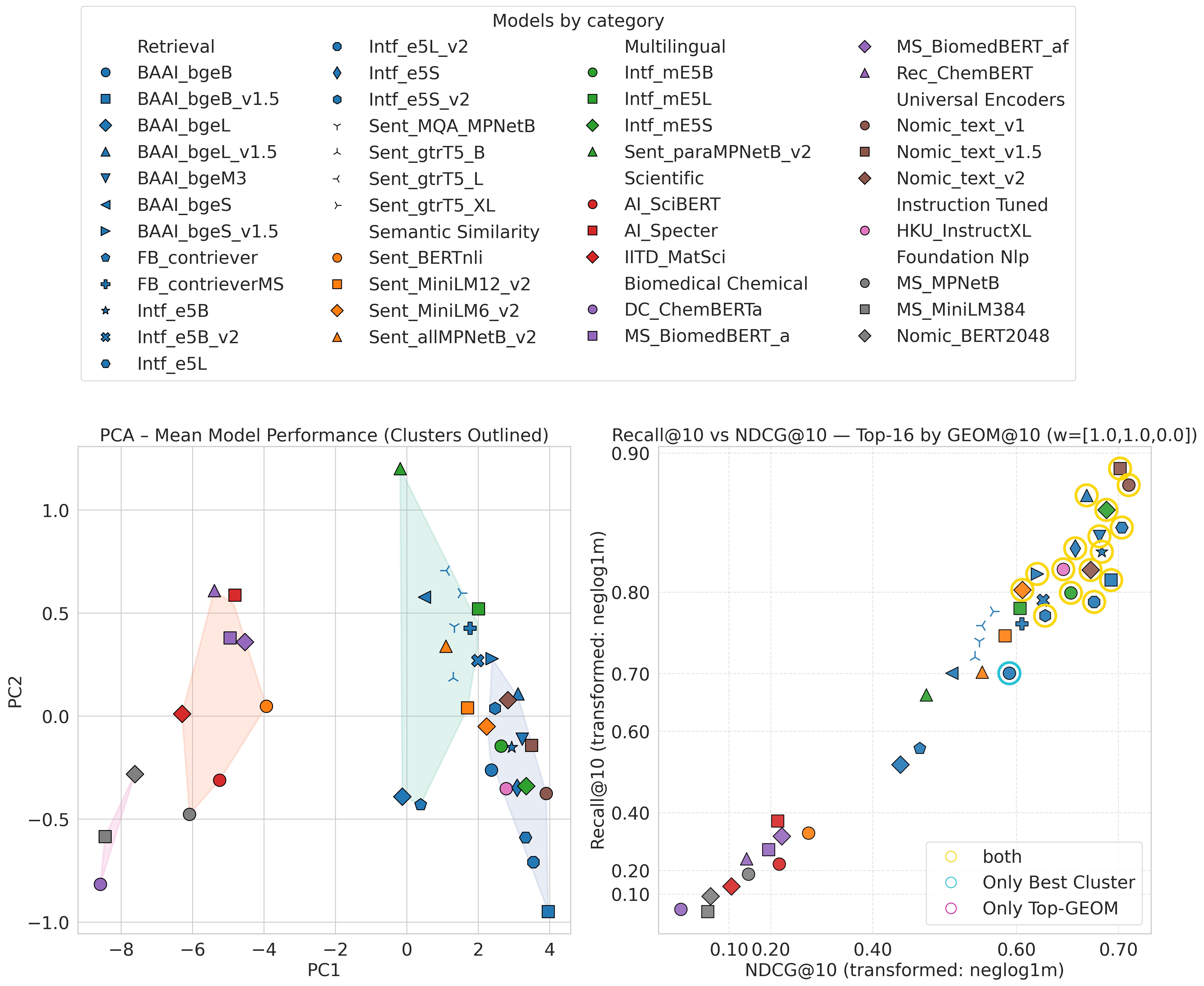}
    \caption{Validation of \(\mathrm{Geom@10}\) as a Phase 1 screening metric. The left panel shows the PCA projection of the 15-metric model profiles with \(k\)-means clusters. The right panel shows \(\mathrm{Recall@10}\) versus \(\mathrm{nDCG@10}\), highlighting the top models by \(\mathrm{Geom@10}\). The high-performing multimetric cluster contains all top-16 models ranked by \(\mathrm{Geom@10}\), showing that the composite metric closely recovers the high-performing region identified by the full metric space.}
    \label{fig:supp_geom10_validation}
\end{figure}

To validate this choice, we represented each model by its full 15-metric retrieval profile. The profile consisted of nDCG, MAP, Recall, Precision, and MRR at cutoffs 1, 10, and 100. We then analyzed the resulting model space using principal component analysis (PCA) followed by \(k\)-means clustering. The first two principal components explained \(99.51\%\) of the total variance (\(\mathrm{PC1}=98.13\%\), \(\mathrm{PC2}=1.38\%\)). Applying \(k\)-means with \(k=4\) identified a cluster whose members occupied the upper-performance region of the PCA projection and the most favorable region of the \(\mathrm{nDCG@10}\)-\(\mathrm{Recall@10}\) plane (Figure~\ref{fig:supp_geom10_validation}).

The high-performing cluster contained 17 models, including all 16 models ranked in the top 16 by \(\mathrm{Geom@10}\). The only additional model in the cluster was {BAAI/bge-base-en}. Thus, the composite metric recovered the high-performing region identified from the full 15-metric retrieval profile with one additional model.

The high-performing cluster consisted primarily of E5, BGE, and Nomic models, together with a small number of other encoders. Its composition was consistent with the per-task rankings.

The external screening, metric-redundancy analysis, and multimetric validation support the use of \(\mathrm{Geom@10}\) for Phase 1 model selection. The resulting 13-model shortlist retains leading retrieval-tuned candidates from both external chemistry benchmarks while keeping the subsequent ChemQuests chunking study computationally tractable.

\subsection{MTEB Task Creation and Qrel Construction from ChemQuests}
\label{sec:supp_mteb_task_creation}

To evaluate retrieval at the chunk level, we converted ChemQuests into a family of local \textsc{MTEB}-compatible retrieval benchmarks. The source material consists of full-text chemistry documents together with question-answer pairs and supporting evidence spans. For each valid chunking configuration, the documents were segmented into chunks, and the resulting chunk collection was used as the retrieval corpus for a configuration-specific benchmark instance. In this way, retrieval performance could be compared across segmentation regimes while keeping the evaluation format fixed.

\begin{algorithm}[H]
\caption{Construction of \textsc{MTEB}-compatible retrieval tasks from ChemQuests}
\label{alg:supp_mteb_task_creation}
\begin{algorithmic}[1]
\Require Full-text documents $\mathcal{D}$, ChemQuests QA pairs $\mathcal{Q}$, chunking configuration $\theta$
\Ensure {corpus.jsonl}, {queries.jsonl}, {qrels.jsonl}

\State $\mathcal{C} \gets \emptyset$
\For{each document $d \in \mathcal{D}$}
    \State $\mathcal{C}_d \gets \textsc{ChunkDocument}(d,\theta)$
    \State $\mathcal{C} \gets \mathcal{C} \cup \mathcal{C}_d$
\EndFor

\For{each chunk $c \in \mathcal{C}$}
    \State write $c$ to {corpus.jsonl}
\EndFor

\For{each QA pair $q \in \mathcal{Q}$}
    \State write $(q\_id,\; q.\text{question})$ to {queries.jsonl}
\EndFor

\For{each QA pair $q \in \mathcal{Q}$}
    \State $d \gets$ source document of $q$
    \State $R \gets q.\text{supporting\_snippet} \;\Vert\; q.\text{question}$
    \State $\mathcal{S} \gets$ chunks from document $d$
    \For{each chunk $c \in \mathcal{S}$}
        \State compute BM25 (Best Matching 25) score between $R$ and $c$
        \State compute lexical similarity score between $R$ and $c$
        \State combine scores into $\mathrm{score}(c)$
    \EndFor
    \State optionally smooth $\mathrm{score}(c)$ with neighboring chunks
    \State mark highest-scoring chunk as relevant
    \State also mark near-tied chunks as relevant
    \State keep at most $K$ relevant chunks
    \State write relevance pairs to {qrels.jsonl}
\EndFor

\State \Return exported benchmark files
\end{algorithmic}
\end{algorithm}

For every configuration, the corresponding benchmark instance was exported in the standard local \textsc{MTEB} retrieval format as three files, {corpus.jsonl}, {queries.jsonl}, and {qrels.jsonl}. The corpus file contains the chunked document units, the query file contains the ChemQuests questions, and the qrels file specifies the chunks labeled as relevant for each query. Here, qrels denotes query relevance labels. At retrieval time, the query text is the question alone. During qrel construction, however, the reference text used for matching was formed by concatenating the supporting evidence snippet with the question. This provides a more stable lexical anchor for locating the evidence-bearing chunk or chunks within the source document.

Relevant chunks were identified using a hybrid lexical matching procedure. For each question, candidate chunks were considered only from the source document associated with that question. This restriction was applied \emph{only during qrel construction} to reduce cross-document noise in the relevance-label construction process. Final retrieval evaluation, by contrast, was always performed over the full chunked corpus for the corresponding configuration.

Each candidate chunk was assigned a relevance score based on three lexical components, namely BM25, Jaccard overlap, and fuzzy token-set similarity. BM25 was computed over the candidate chunk set and then normalized by the maximum BM25 score observed for that query. Jaccard overlap and fuzzy similarity were combined linearly to form a lexical-overlap term. To account for evidence that may extend across chunk boundaries, this lexical term was optionally smoothed using a neighbor-aware mechanism with an adjacency window of one chunk on each side. In this study, the final lexical score is a convex combination of the local chunk score and the score of the merged neighborhood, with neighbor interpolation weight $\alpha = 0.2$.

The final score assigned to a candidate chunk $c$ is
\[
s(c)=w_{\mathrm{BM25}}\hat{b}(c)+\tilde{\ell}(c),
\]
where $\hat{b}(c)$ denotes the BM25 score normalized by the maximum BM25 value for the query, and $\tilde{\ell}(c)$ denotes the neighbor-aware lexical similarity term derived from Jaccard and fuzzy token-set overlap. In the present study, all component weights were set to 1.0.

Relevant chunks were selected using a relative-threshold rule. For each query, the highest-scoring chunk was always retained as relevant. Additional chunks were also retained when their score was at least 95\% of the best score for that query. To avoid overly diffuse supervision, the number of retained chunks per query was capped at five. The resulting qrels therefore provide binary relevance labels anchored to the strongest local evidence in the source document, while still allowing multiple nearby chunks to be labeled relevant when the answer evidence spans chunk boundaries.

This benchmark-construction procedure yields one local \textsc{MTEB}-compatible retrieval task for each valid chunked version of the ChemQuests corpus. It therefore enables controlled evaluation of how chunking decisions affect retrieval quality while keeping the query set and relevance-labeling procedure consistent across configurations.

\subsection{Evaluation Pipeline Algorithms}
\label{sec:supp_evaluation_pipeline_algorithms}

Algorithms~\ref{alg:supp_phase1_screening} and~\ref{alg:supp_phase2_evaluation} summarize the two-stage evaluation pipeline used for model screening and ChemQuests retrieval evaluation.

\begin{algorithm}[t]
\caption{Phase 1: Screening and selecting embedding models}
\label{alg:supp_phase1_screening}
\begin{algorithmic}[1]
\Require Candidate embedding models $\mathcal{M}$, screening tasks $\mathcal{T}=\{{ChemNQRetrieval}, {ChemHotpotQARetrieval}\}$, retention ratio $p=0.2$
\Ensure Shortlisted model set $\mathcal{M}^{\star}$

\For{each model $m \in \mathcal{M}$}
    \For{each screening task $t \in \mathcal{T}$}
        \State evaluate $m$ on $t$ using MTEB
        \State save raw result file for $(m,t)$
    \EndFor
\EndFor

\State $\mathcal{M}^{\star} \gets \emptyset$

\For{each task $t \in \mathcal{T}$}
    \State load all saved results for task $t$
    \For{each model $m$ with a valid result on $t$}
        \State choose the first available split in the order:
        \Statex \hspace{\algorithmicindent} {test} $\rightarrow$ {validation} $\rightarrow$ {train}
        \State extract \(\mathrm{nDCG@10}\) and \(\mathrm{Recall@10}\)
        \State compute $\mathrm{Geom@10}(m,t)$
    \EndFor
    \State rank models by $\mathrm{Geom@10}$ on task $t$
    \State retain the top $p$ fraction of models for task $t$
    \State $\mathcal{M}^{\star} \gets \mathcal{M}^{\star} \cup \{\text{retained models on } t\}$
\EndFor

\State save per-task retained models and the union shortlist
\State \Return $\mathcal{M}^{\star}$
\end{algorithmic}
\end{algorithm}

\begin{algorithm}[t]
\caption{Phase 2: Evaluating shortlisted models across chunking configurations on ChemQuests}
\label{alg:supp_phase2_evaluation}
\begin{algorithmic}[1]
\Require Shortlisted models $\mathcal{M}^{\star}$, ChemQuests repository $\mathcal{R}$, chunking strategies $\mathcal{H}$, chunk sizes $\mathcal{L}$, overlaps $\mathcal{O}$
\Ensure Retrieval results for all valid model-chunking configurations

\State download ChemQuests files from $\mathcal{R}$:
\Statex \hspace{\algorithmicindent} {metadata.jsonl}, {full\_text.jsonl}, {qa.jsonl}

\For{each model $m \in \mathcal{M}^{\star}$}
    \For{each chunking strategy $h \in \mathcal{H}$}
        \For{each chunk size $L \in \mathcal{L}$}
            \For{each overlap $O \in \mathcal{O}$}
                \If{$O > \min(64,0.3L)$}
                    \State skip configuration
                \ElsIf{$L-O < 96$}
                    \State skip configuration
                \ElsIf{$L+O > 512$}
                    \State skip configuration
                \Else
                    \State instantiate configuration for $(m,h,L,O)$
                    \State chunk {full\_text.jsonl} using chunking strategy $h$ with $(L,O)$
                    \State save resulting chunks to {chunks.jsonl}
                    \State build a local \textsc{MTEB}-compatible ChemQuests task from:
                    \Statex \hspace{\algorithmicindent} {qa.jsonl} and {chunks.jsonl}
                    \Statex \hspace{\algorithmicindent} using Algorithm~\ref{alg:supp_mteb_task_creation}
                    \State evaluate model $m$ on the generated ChemQuests retrieval task using MTEB
                    \State save task outputs and summary metrics
                \EndIf
            \EndFor
        \EndFor
    \EndFor
\EndFor

\State \Return all saved retrieval results
\end{algorithmic}
\end{algorithm}

\subsection{Evaluation Time Analysis}
\label{sec:supp_eval_time_analysis}

We examined computational efficiency using evaluation time as the principal cost indicator. In Phase 2, the total duration of each evaluated configuration was decomposed into document segmentation, local benchmark construction, and retrieval evaluation on the derived \textsc{MTEB}-compatible task. This decomposition identifies which stages contribute to variation in total evaluation time.

\begin{figure}[!htbp]
    \centering
    \includegraphics[width=0.9\textwidth]{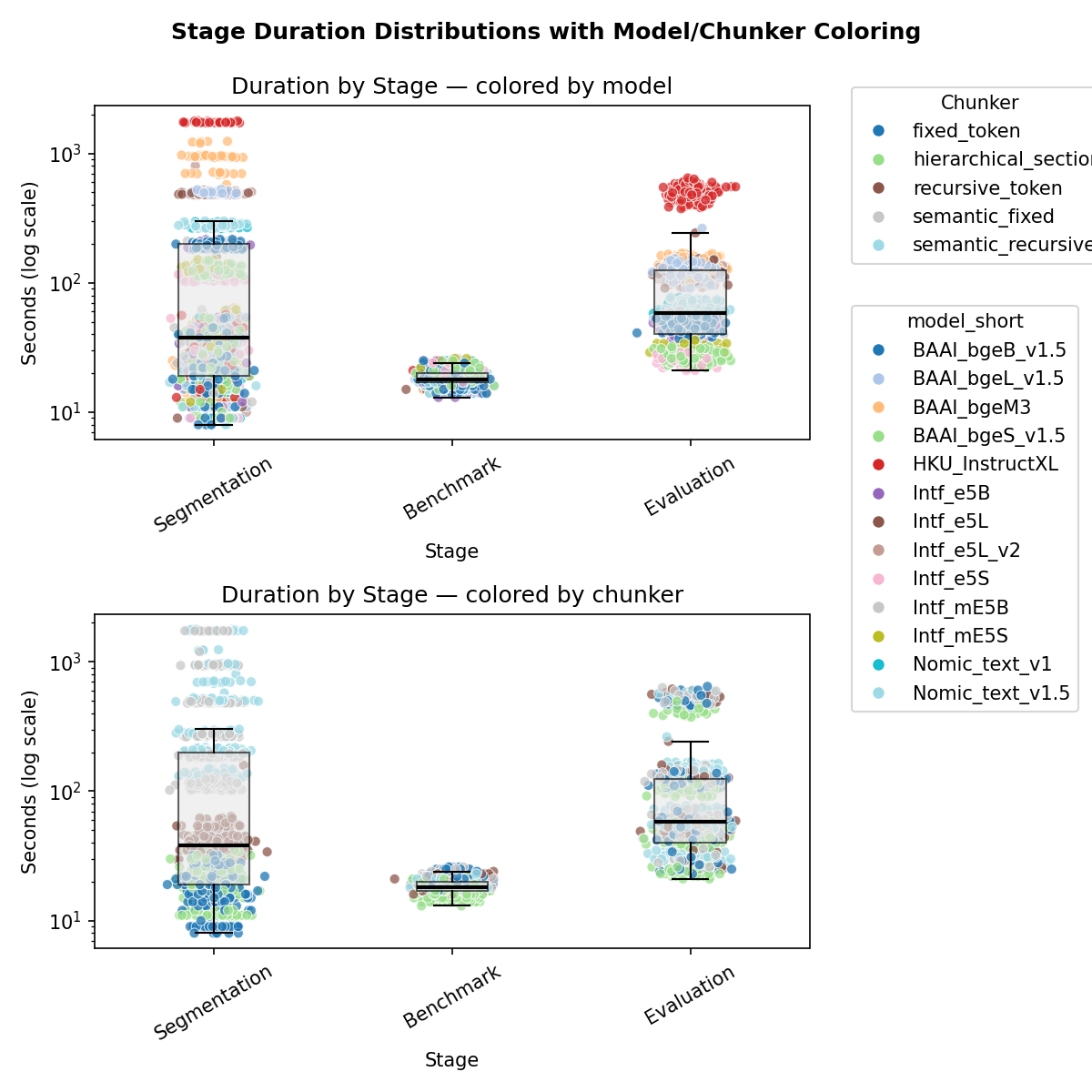}
    \caption{Stage-wise evaluation-time distributions for the three main Phase 2 pipeline stages: Segmentation, Benchmark, and Evaluation.
    Evaluation times are shown separately for Segmentation (document segmentation), Benchmark (construction of the local \textsc{MTEB}-compatible retrieval benchmark), and Evaluation (retrieval evaluation of the embedding model on the derived task). The top panel colors runs by embedding model, and the bottom panel colors runs by chunking strategy; the y-axis is shown on a logarithmic scale. Benchmark construction is consistently short and stable, whereas most variation in evaluation time arises from document segmentation and retrieval evaluation. Differences among chunking strategies are most visible during document segmentation, while differences among models are most pronounced during retrieval evaluation.}
    \label{fig:supp_eval_time_stage_distributions}
\end{figure}

Figure~\ref{fig:supp_eval_time_stage_distributions} shows the distributions of evaluation times for these three components across all valid Phase 2 configurations. Benchmark construction is short and comparatively stable across runs and therefore contributes little to variation in total evaluation time. Most of the variation arises from document segmentation and retrieval evaluation. The panel grouped by chunking strategy shows that semantic chunking generally requires more segmentation time than fixed-token, recursive-token, and hierarchical-section chunking. The panel grouped by embedding model shows that retrieval-evaluation time varies among models.

\begin{figure}[!htbp]
    \centering
    \includegraphics[width=0.85\textwidth]{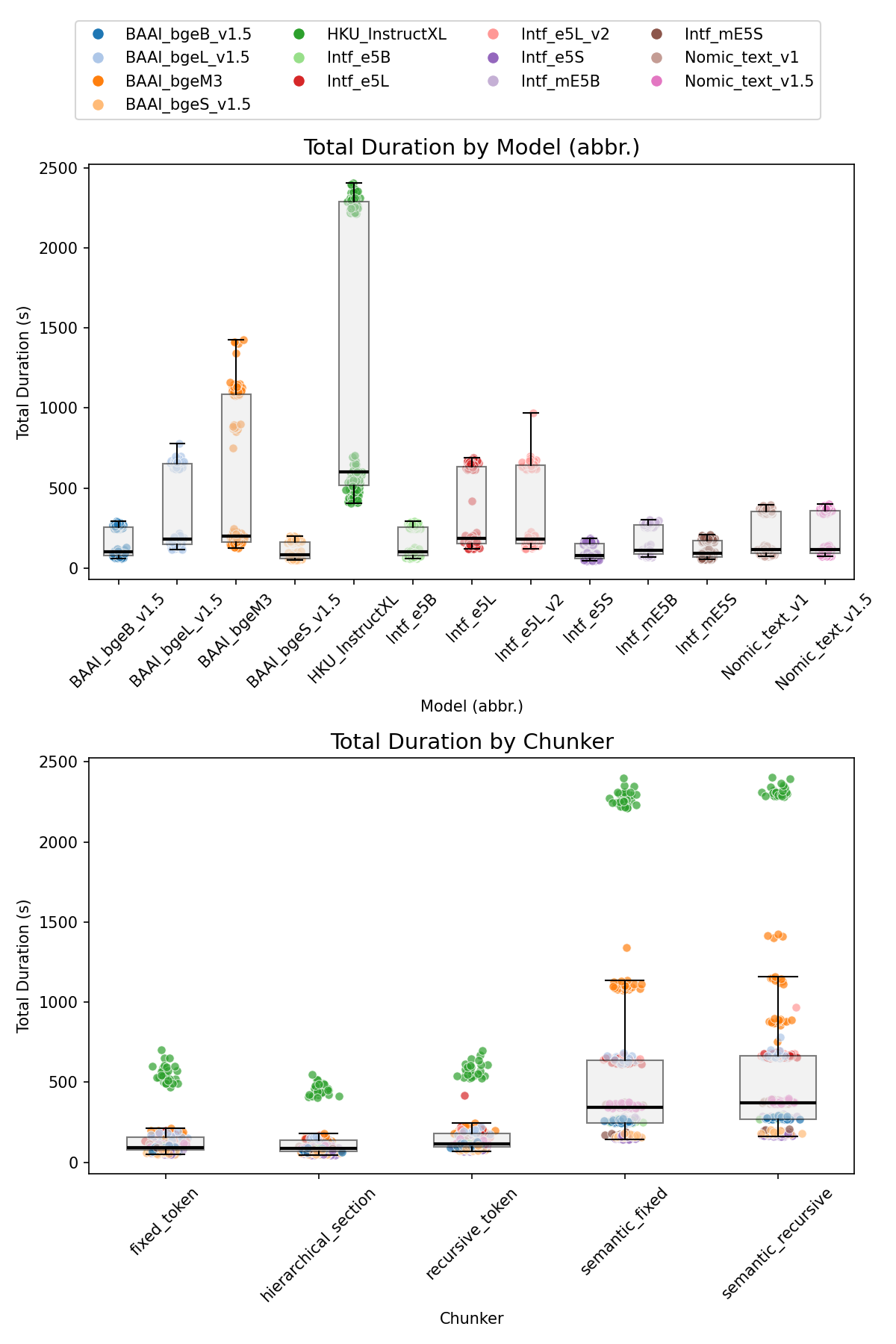}
    \caption{Total evaluation time across Phase 2 configurations grouped by major design variables.
    Boxplots summarize the distribution of total evaluation time for all valid configurations when grouped by embedding model and chunking strategy. Embedding-model choice produces the largest observed differences in evaluation time. Chunking strategy has the next largest association, with semantic chunking generally slower than fixed-token, recursive-token, and hierarchical-section chunking.}
    \label{fig:supp_eval_time_boxplots}
\end{figure}

This decomposition is reflected in the aggregate summaries of total evaluation time in Figure~\ref{fig:supp_eval_time_boxplots}. Embedding-model choice is associated with the largest differences in total evaluation time. The model-wise boxplots display larger differences in central tendency and spread than the corresponding boxplots for chunk size or overlap.

\begin{figure}[!htbp]
    \centering
    \includegraphics[width=0.90\textwidth]{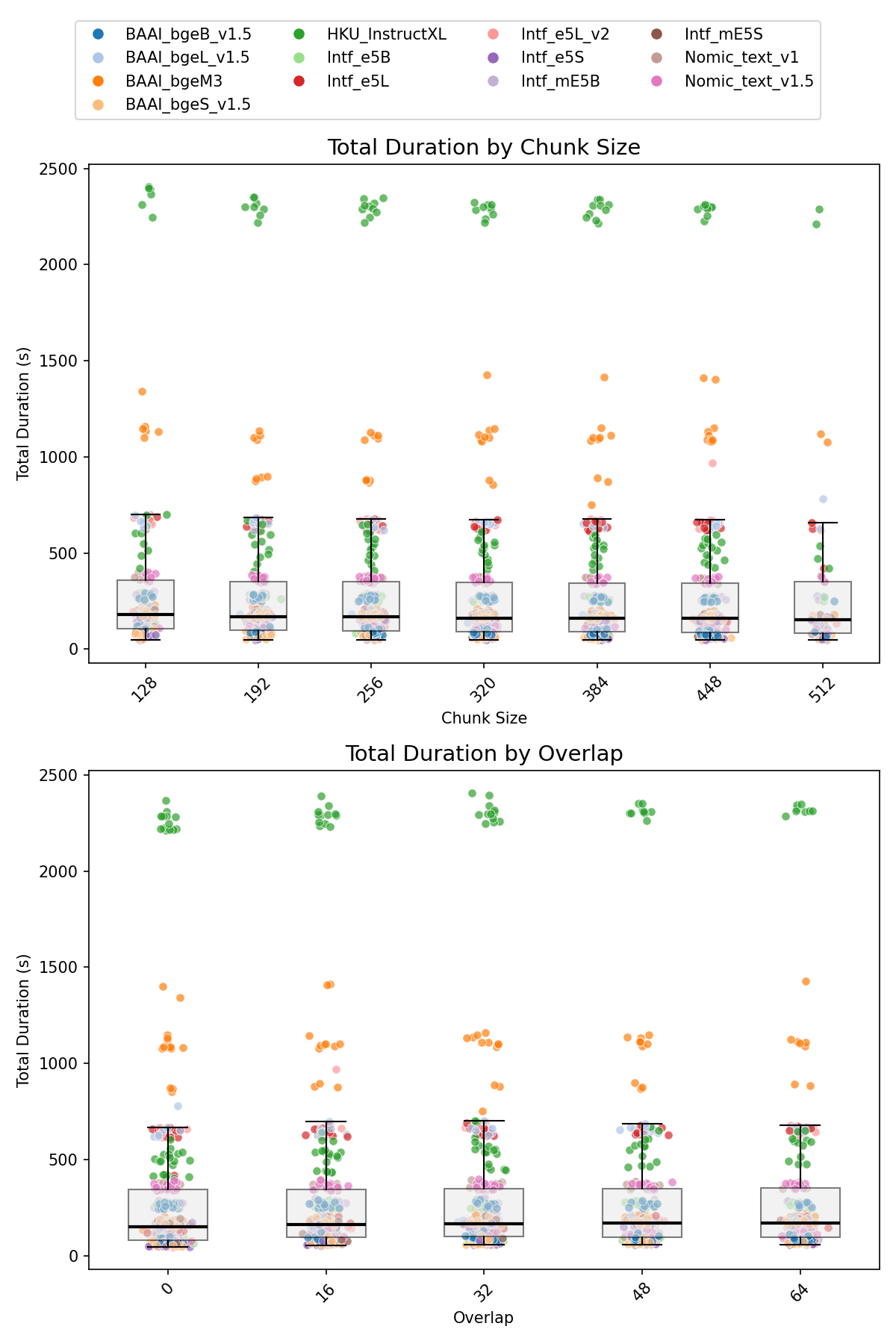}
    \caption{Total evaluation time across Phase 2 configurations grouped by major design variables.
    Boxplots summarize the distribution of total evaluation time for all valid configurations when grouped by chunk size and overlap. Chunk size and overlap show smaller but still visible associations, with broadly similar distributions across settings compared with the variation associated with model choice.}
    \label{fig:supp_eval_time_boxplots_cont}
\end{figure}

Chunking strategy shows the second-largest association with evaluation time. As shown in Figure~\ref{fig:supp_eval_time_boxplots}, semantic chunking is consistently slower than the other evaluated strategies. In particular, {semantic-fixed} and {semantic-recursive} tend to occupy the highest evaluation-time range, whereas {fixed-token} and {hierarchical-section} consistently occupy the lower range, with {recursive-token} generally intermediate.

Chunk size and overlap have smaller but still visible associations with total evaluation time. Figure~\ref{fig:supp_eval_time_boxplots_cont} shows broadly similar evaluation-time distributions across chunk sizes and overlap settings, especially when compared with the larger differences among models. Smaller chunks and larger overlaps tend to produce more indexed units and therefore require more work during segmentation and evaluation. These associations remain smaller than those for embedding model and chunking strategy. Because the evaluated grid is constrained and does not contain all chunk-size--overlap combinations, these trends should be interpreted as directional rather than as a balanced factorial comparison.

Most observed variation in evaluation time is associated with embedding-model choice and, secondarily, chunking strategy. Benchmark construction contributes little to total evaluation time, chunk size and overlap have smaller associations, and semantic chunking requires more segmentation time than the other evaluated strategies.

\subsection{Top-Performing Configurations Across Metrics}
\label{sec:supp_top_configs_across_metrics}

\begin{figure}[!htbp]
    \centering
    \includegraphics[width=0.82\textwidth,height=0.72\textheight,keepaspectratio]{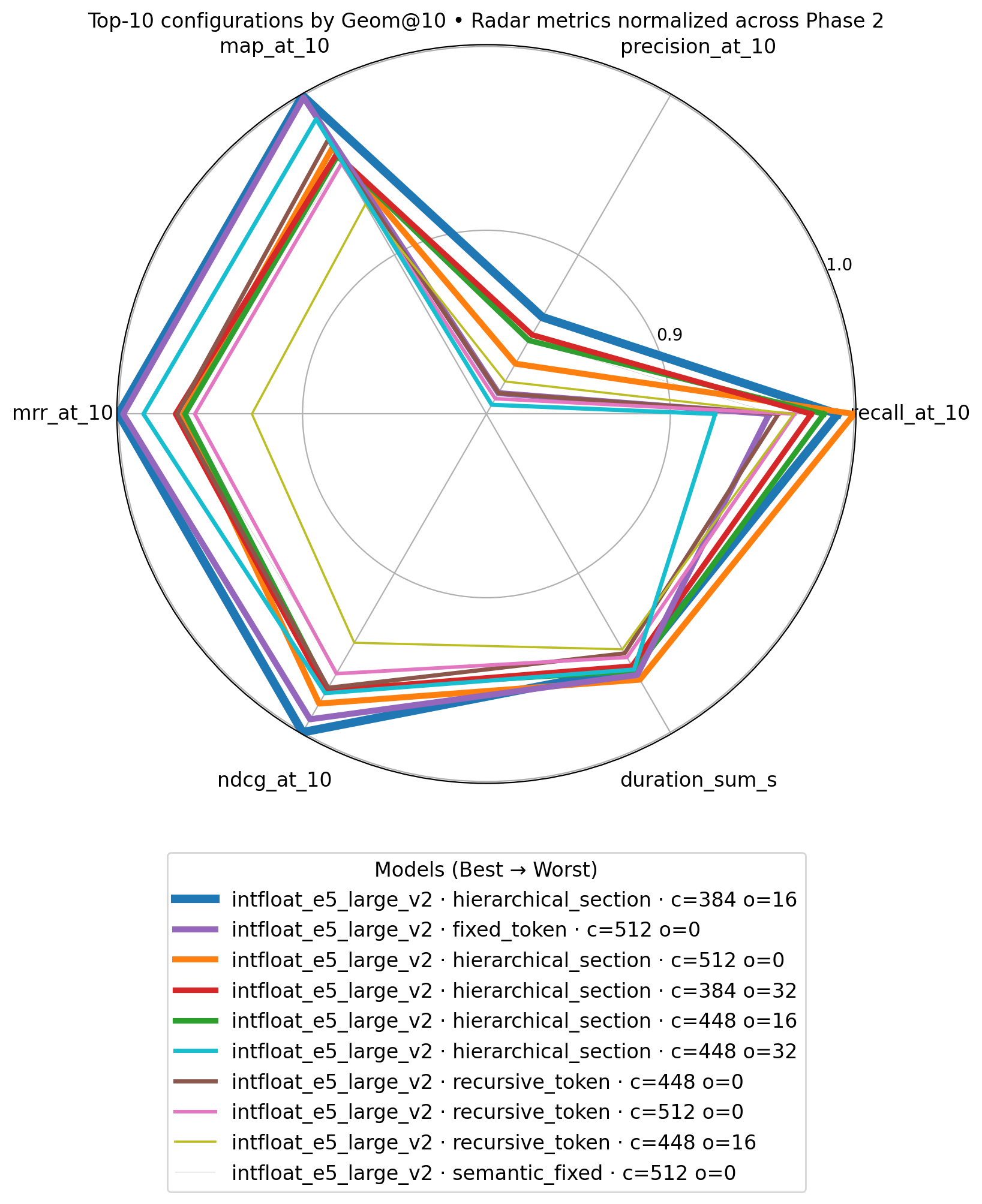}
    \caption{Multimetric comparison of top-performing configurations. Radar plot of the 10 highest-ranked configurations selected from the top \(1\%\) of all runs by \(\mathrm{Geom@10}\). The displayed axes are normalized \(\mathrm{Recall@10}\), \(\mathrm{Precision@10}\), \(\mathrm{MAP@10}\), \(\mathrm{MRR@10}\), \(\mathrm{nDCG@10}\), and evaluation time; evaluation time is inverted during normalization so that larger radial extent corresponds to faster evaluation. All ten selected configurations use {intfloat/e5-large-v2}. They use 384--512-token chunks and overlaps of 0--32 tokens, and are predominantly paired with fixed-token, recursive-token, or hierarchical-section chunking.}
    \label{fig:supp_top_configs_across_metrics}
\end{figure}

We compared the top 10 configurations by \(\mathrm{Geom@10}\), all of which belonged to the top \(1\%\) of runs, across the principal retrieval metrics and evaluation time using a normalized radar plot (Figure~\ref{fig:supp_top_configs_across_metrics}). The axes show normalized \(\mathrm{Recall@10}\), \(\mathrm{Precision@10}\), \(\mathrm{MAP@10}\), \(\mathrm{MRR@10}\), \(\mathrm{nDCG@10}\), and evaluation time. Evaluation time is inverted during normalization so that a larger radial extent represents faster evaluation.

All ten selected configurations use {intfloat/e5-large-v2}.

The selected configurations have high normalized values for \(\mathrm{nDCG@10}\), \(\mathrm{MAP@10}\), \(\mathrm{MRR@10}\), and \(\mathrm{Recall@10}\), as well as similar values for \(\mathrm{Precision@10}\).

Most selected configurations use chunk sizes of 448 or 512 tokens. All use overlaps of at most 32 tokens, with five of the ten using zero overlap. Fixed-token, recursive-token, and hierarchical-section chunking occur most frequently, while semantic chunking occurs less frequently.

All selected configurations use {intfloat/e5-large-v2} and show high values across the displayed retrieval metrics. Several chunking variants paired with this model are also similar in their normalized profiles. In particular, the fixed-token, hierarchical-section, and recursive-token variants with 512-token chunks and zero overlap lie near the outer boundary on most axes. Their differences are small relative to the separation from the remaining configurations.

\begin{figure}[!htbp]
    \centering
    \includegraphics[width=\textwidth,height=0.72\textheight,keepaspectratio]{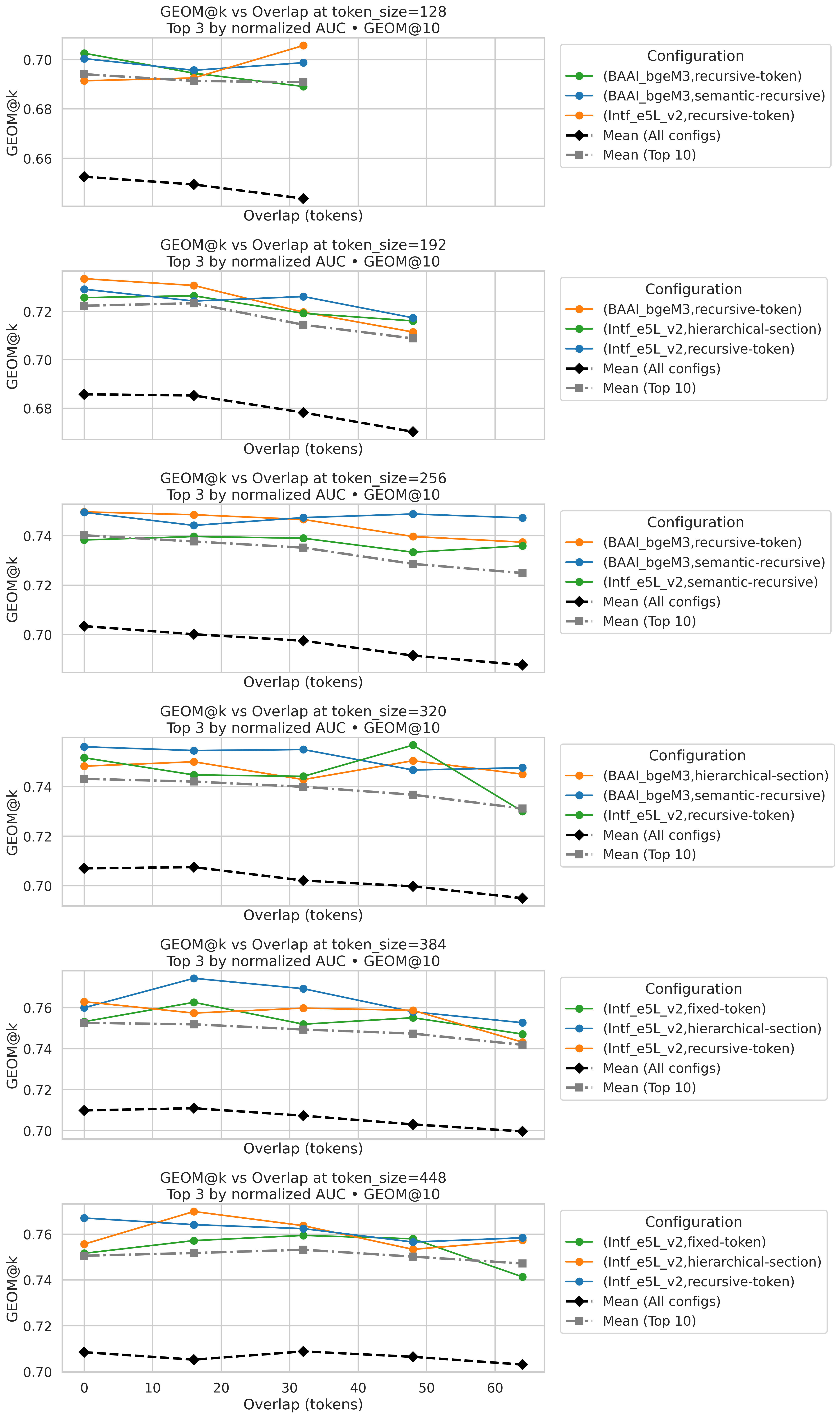}
    \caption{Effect of overlap on retrieval effectiveness across fixed chunk sizes. Curves show \(\mathrm{Geom@10}\) as a function of chunk overlap for selected top-performing configurations, together with the mean over all configurations and the mean over the top-performing subset, evaluated separately for each fixed chunk size. The highlighted configurations are selected by normalized area under the curve (AUC) within each size-specific overlap sweep. Across most chunk sizes included in this analysis, average retrieval performance declines as overlap increases, although some top configurations exhibit small local optima at low nonzero overlap values such as 16 or 32 tokens. Because the constrained grid did not provide sufficient overlap variation for 512-token chunks, this figure should not be read as an overlap analysis at exactly 512 tokens.}
    \label{fig:supp_overlap_auc_curves}
\end{figure}

\begin{figure}[!htbp]
    \centering
    \includegraphics[width=\textwidth,height=0.72\textheight,keepaspectratio]{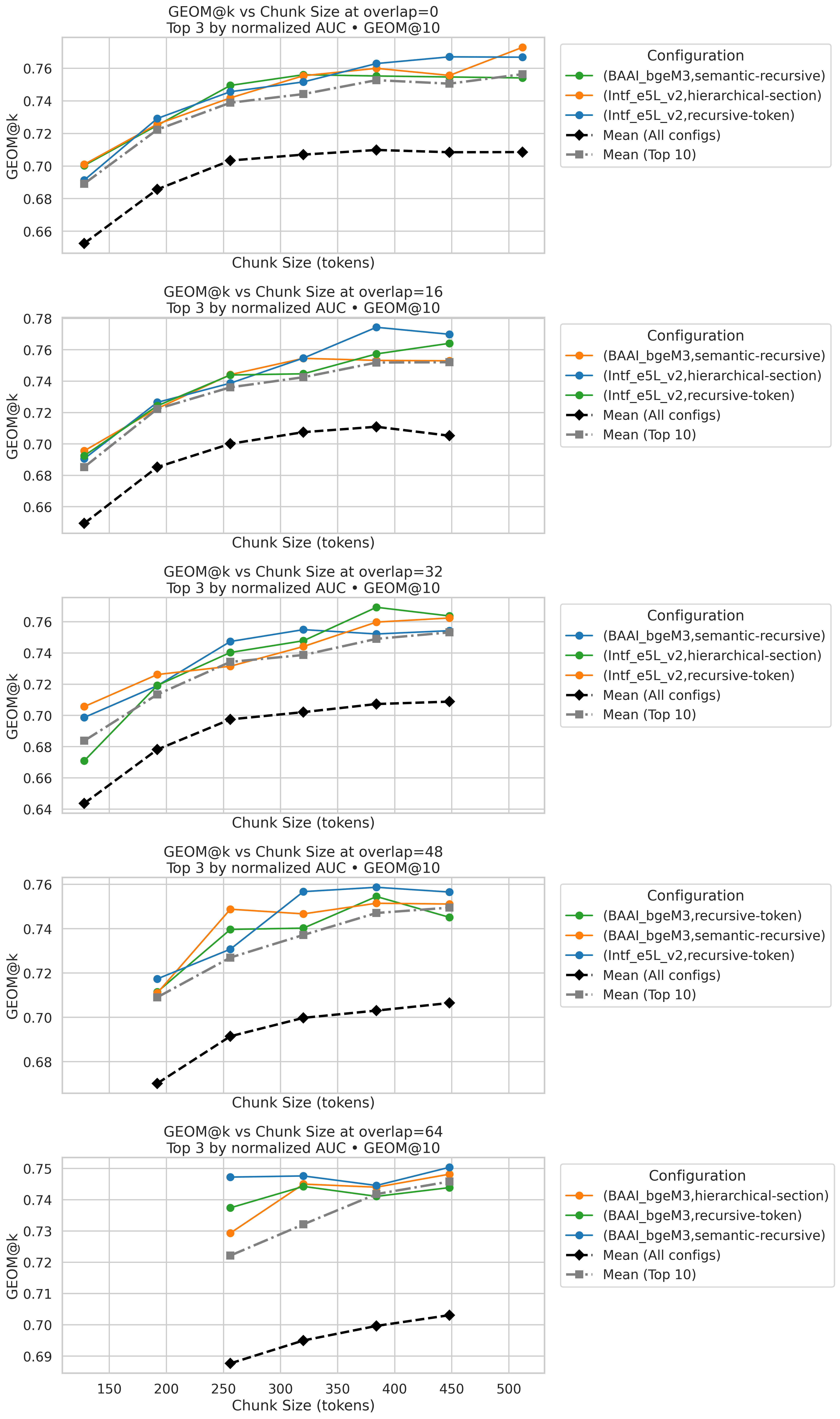}
    \caption{Effect of chunk size on retrieval effectiveness across fixed overlap settings. Curves show \(\mathrm{Geom@10}\) as a function of chunk size for selected top-performing configurations, together with the mean over all configurations and the mean over the top-performing subset, evaluated separately for each fixed overlap value. The highlighted configurations are selected by normalized area under the curve (AUC) within each overlap-specific chunk-size sweep. Retrieval effectiveness generally increases from small chunk sizes toward medium-to-large sizes, with gains tending to level off in the larger-size regime. In particular, 448-token chunks remain competitive across the nonzero-overlap curves, while 512-token chunks perform strongly where evaluated, especially at zero overlap.}
    \label{fig:supp_chunksize_auc_curves}
\end{figure}
\FloatBarrier

\clearpage
\bibliographystyle{unsrtnat}
\bibliography{references}

\end{document}